\documentclass{bmcart}

\usepackage[utf8]{inputenc} %unicode support
\startlocaldefs
\usepackage{url}
\usepackage{graphicx}
\usepackage{booktabs}
\usepackage{multirow}
\usepackage{multicol}
\usepackage[caption=false]{subfig}
\usepackage{float}
\usepackage{amsmath}
\usepackage{amsfonts}
\usepackage{diagbox}
\usepackage{enumitem}
\usepackage{enotez}
\setenotez{counter-format=alph,list-name=Endnotes}
\let\footnote=\endnote
\usepackage{lineno}
\modulolinenumbers[2]

\DeclareMathOperator*{\argmax}{arg\,max}

\newcommand{\rev}[1]{\textcolor{black}{#1}}

\endlocaldefs

\begin{document}

%%% Start of article front matter
\begin{frontmatter}

\begin{fmbox}
\dochead{Research}

\title{Support the Underground: Characteristics of Beyond-Mainstream Music Listeners}

\author[
   addressref={aff1},                   
   email={dkowald@know-center.at}  
]{\inits{DK}\fnm{Dominik} \snm{Kowald}}
\author[
   addressref={aff1},
   email={pmuellner@know-center.at}
]{\inits{PM}\fnm{Peter} \snm{Muellner}}
\author[
   addressref={aff2},
   email={eva.zangerle@uibk.ac.at}
]{\inits{EZ}\fnm{Eva} \snm{Zangerle}}
\author[
   addressref={aff3},
   email={c.bauer@uu.nl}
]{\inits{CB}\fnm{Christine} \snm{Bauer}}
\author[
   addressref={aff4,aff5},
   email={markus.schedl@jku.at}
]{\inits{MS}\fnm{Markus} \snm{Schedl}}
\author[
   addressref={aff6},
   email={elisabeth.lex@tugraz.at}
]{\inits{EL}\fnm{Elisabeth} \snm{Lex}}

\address[id=aff1]{%                   
  \orgname{Know-Center GmbH}, 
  \city{Graz},                           
  \cny{Austria}                  
}
\address[id=aff2]{%
  \orgname{University of Innsbruck},
  \city{Innsbruck},
  \cny{Austria}
}
\address[id=aff3]{%
  \orgname{Utrecht University},
  \city{Utrecht},
  \cny{The Netherlands}
  }
\address[id=aff4]{%
  \orgname{Johannes Kepler University Linz},
  \city{Linz},
  \cny{Austria}
}
\address[id=aff5]{%
  \orgname{Linz Institute of Technology AI Lab},
  \city{Linz},
  \cny{Austria}
}
\address[id=aff6]{%
  \orgname{Graz University of Technology},
  \city{Graz},
  \cny{Austria}
}
%%%%%%%%%%%%%%%%%%%%%%%%%%%%%%%%%%%%%%%%%%%%%
%\begin{artnotes}
%\note{Sample of title note}     % note to the article
%\note[id=n1]{Equal contributor} % note, connected to author
%\end{artnotes}

\end{fmbox}% comment this for two column layout

%%%%%%%%%%%%%%%%%%%%%%%%%%%%%%%%%%%%%%%%%%%%%%

\begin{abstractbox}

\begin{abstract}
Music recommender systems have become an integral part of music streaming services such as Spotify and Last.fm to assist users navigating the extensive music collections offered by them. However, while music listeners interested in mainstream music are traditionally served well by music recommender systems, users interested in music beyond the mainstream (i.e., non-popular music) rarely receive relevant recommendations. In this paper, we study the characteristics of beyond-mainstream \rev{music and} music listeners and analyze to what extent these characteristics impact the quality of music recommendations provided. Therefore, we create a novel dataset consisting of Last.fm listening histories of several thousand beyond-mainstream music listeners, which we enrich with additional metadata describing music tracks and music listeners. \rev{Our analysis of this dataset shows} four subgroups within the group of beyond-mainstream music listeners that differ not only with respect to their preferred music but also with their demographic characteristics. Furthermore, we evaluate the quality of music recommendations that these subgroups are provided with four different recommendation algorithms where we find significant differences between the groups. Specifically, our results show a positive correlation between a subgroup's openness towards music listened to by members of other subgroups and recommendation accuracy. We believe that our findings provide valuable insights for developing improved user models and recommendation approaches to better serve beyond-mainstream music listeners.
\end{abstract}


\end{abstractbox}
%
%\end{fmbox}% uncomment this for twcolumn layout

\end{frontmatter}

%%%%%%%%%%%%%%%%%%%%%%%%%%%%%%%%%%%%%%%%%%%%%%%%%%%%%%%%%%%%%%%%%%%%%%%%%%%%%%%%

%\begin{linenumbers}
\section{Introduction}
\label{s:intro}
In the digital era, users have access to continually increasing amounts of music via music streaming services such as Spotify and Last.fm. Music recommender systems have become an essential means to help users deal with content and choice overload as they assist users in searching, sorting, and filtering these extensive music collections. Simultaneously, both music listeners and artists benefit from the employed segmentation and personalization approaches that are typically leveraged in music recommendation approaches~\cite{schedl2015music}. As a result, users with different preferences and needs can be targeted in various ways with the goal that all users are presented the information and content that they need or prefer.
\rev{This also means that current recommendation techniques should serve all users equally well, independent of their inclination to popular content.}

\begin{figure}[t!]
    \centering
    \includegraphics[width=0.70\textwidth]{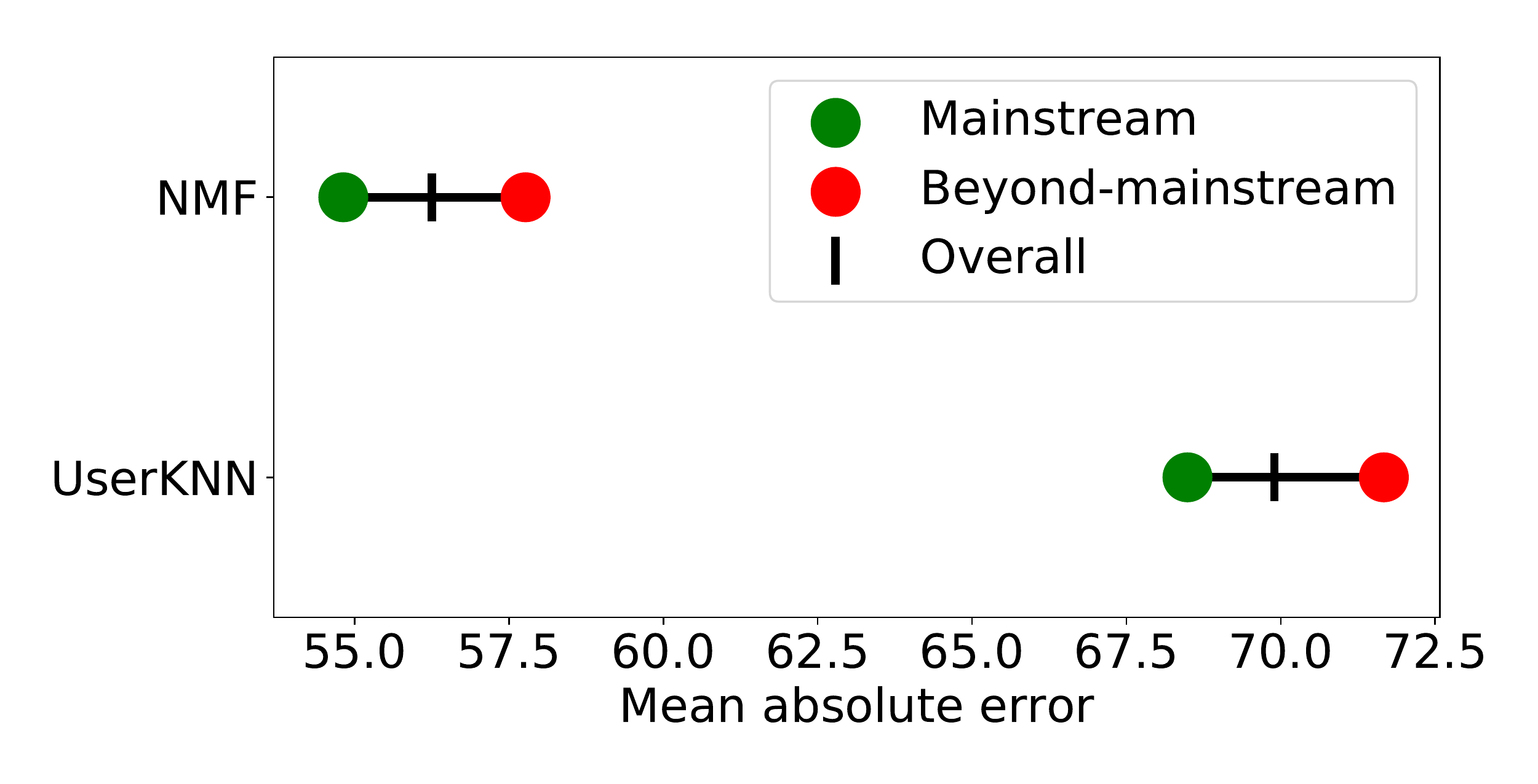}
    \caption{Recommendation accuracy measured by the mean absolute error (MAE) of a non-negative matrix factorization-based approach (i.e., NMF~\cite{luo2014efficient}) and a neighborhood-based approach (i.e., UserKNN~\cite{herlocker2004evaluating}) for mainstream and beyond-mainstream user groups in Last.fm. We see that beyond-mainstream users receive a substantially lower recommendation quality (i.e., higher MAE) compared to mainstream music listeners.
    \rev{Thus, for recommender systems, it is harder to provide high-quality recommendations to beyond-mainstream music listeners than to mainstream music listeners.}}
     \label{fig:motivation_example}
\end{figure}

\paragraph{\textbf{Present work.}} In the paper at hand, we focus on music consumers \rev{who} listen to music beyond the mainstream (i.e., users who listen to non-popular music) in the music streaming platform Last.fm\footnote{\url{https://www.last.fm/}}. As highlighted in Figure~\ref{fig:motivation_example}, current recommender systems do not work well for consumers of beyond-mainstream music (see Section~\ref{s:rec_first} for details on this analysis). In contrast, music consumers who listen to popular music seem to get better recommendations. 
\rev{This finding is not essentially new. In fact, it is a widely-known problem that recommender systems (and those based on collaborative filtering, in particular) are prone to popularity bias, which leads to the behavior that long-tail items (i.e., items with few user interactions) have little chance being recommended. This phenomenon is also present across different application domains such as movies~\cite{abdollahpouri2019unfairness} or music~\cite{celma2009music}}.

Our previous work~\cite{kowald2020unfairness} has shown that users interested in beyond-mainstream music tend to have larger user profile sizes \rev{(i.e., individual users show a high(er) number of distinct artists they have listened to)} compared to users interested in mainstream music. The observation that beyond-mainstream music listeners produce a substantial amount of digital footprints motivates the need to improve the recommendation quality for this group.
\rev{However, although related research has already studied the long-tail recommendation problem (e.g.,~\cite{celma2008hits,celma:springer:2010,Oord2013:DCM,goel2010anatomy}; see Section~\ref{s:related} for a more detailed discussion of related work), it is still a fundamental challenge to understand and identify the characteristics of beyond-mainstream music and beyond-mainstream music listeners. 
Additionally, related work~\cite{tintarev2013adapting} has shown that the group-specific concepts of openness and diversity influence recommendation quality, where openness is defined as across-group diversity (i.e., do users of one group listen to the music of other groups?) and diversity is defined as within-group variability (i.e., how dissimilar is the music listened to by users within groups?). Thus, we are also interested in the correlation between the characteristics of beyond-mainstream music and music listeners with openness and diversity patterns as well as with recommendation quality. Concretely, our work is guided by the following research question:}

\rev{\emph{RQ: What are the characteristics of beyond-mainstream music tracks and music listeners, and how do these characteristics correlate with openness and diversity patterns as well as with recommendation quality?}}

To address this research question, we create, provide, and analyze a novel dataset \rev{called} \emph{LFM-BeyMS}, which contains complete listening histories of more than 2,000 beyond-mainstream music listeners mined from the Last.fm music streaming platform. Besides, our dataset is enriched with acoustic features and genres of music tracks. 
\rev{Using this enriched dataset, we identify different types of beyond-mainstream music via unsupervised clustering applied to the acoustic features of music tracks. We then characterize the resulting music clusters using music genres. Then, we assign beyond-mainstream users to the clusters to further divide the beyond-mainstream users into subgroups. We study how the characteristics of these beyond-mainstream subgroups correlate with openness and diversity patterns as well as with recommendation quality measured through prediction accuracy.} 

\paragraph{\textbf{Findings and contributions.}} \rev{We identify four clusters of beyond-mainstream music in our dataset: (i)~$C_{folk}$, music with high acousticness such as ``folk'', (ii)~$C_{hard}$, high energy music such as ``hardrock'', (iii)~$C_{ambi}$, music with high acousticness and high instrumentalness such as ``ambient'', and (iv)~$C_{elec}$, music with high energy and high instrumentalness such as ``electronica''. 
By assigning users to these clusters, we get four distinct subgroups of beyond-mainstream music listeners: (i)~$U_{folk}$, (ii)~$U_{hard}$, (iii)~$U_{ambi}$, and (iv)~$U_{elec}$.} 
We also find that these groups differ considerably with respect to the accuracy of recommendations they receive, where group $U_{ambi}$ gets significantly better recommendations than $U_{hard}$. 
\rev{When relating our results to openness and diversity patterns of the subgroups, we find that $U_{ambi}$ is the most open but least diverse group, while $U_{hard}$ is the least open but most diverse group. This is in line with related research~\cite{tintarev2013adapting}, which has shown that openness is stronger correlated with accurate recommendations than diversity. This means that users are more likely to accept recommendations from different groups (i.e., openness) rather than varied within a group (i.e., diversity).}

Summed up, our contributions are:
\begin{itemize}
    \item We identify more than 2,000 beyond-mainstream music listeners on the Last.fm platform and enrich their \rev{listening profiles} with acoustic features and genres of music tracks listened to (Sections~\ref{sec:acoustic_features}--\ref{s:beyond-music}).
    \item We \rev{validate related research by showing} that beyond-mainstream music listeners receive a significantly lower recommendation accuracy than mainstream music listeners (Section~\ref{s:rec_first}).  
    \item \rev{We identify four clusters of beyond-mainstream music using unsupervised clustering and characterize them with respect to acoustic features and music genres (Section~\ref{s:clustering}).}
    \item \rev{We define four subgroups of beyond-mainstream music listeners by assigning users to the music clusters and discuss the relationship between openness, diversity, and recommendation quality of these groups (Section~\ref{subsec:user_clustering}).}
    \item To foster reproducibility of our research, we make available our novel \emph{LFM-BeyMS} dataset via Zenodo\footnote{\url{https://doi.org/10.5281/zenodo.3784764}} and the entire Python-based implementation of our analyses via Github\footnote{\url{https://github.com/pmuellner/supporttheunderground}}. 
\end{itemize}
We believe that our findings provide useful insights for creating user models and recommendation algorithms that better serve beyond-mainstream music listeners.

\section{Related Work}
\label{s:related}
We identify \rev{three} strands of research that are relevant to our work: (i) modeling of music preferences, \rev{(ii) long-tail recommendations,} and (iii) \rev{popularity bias in} music recommender systems.

\paragraph{\textbf{Modeling of music preferences.}} A multitude of factors~\cite{schedl2018current} influences musical tastes and musical preferences of users. Characteristics of music listeners and music preferences have been studied in various research domains~\cite{haas2010music}, ranging from music sociology~\cite{adorno1988introduction} and psychology~\cite{deutsch2013psychology} to music information retrieval and music recommender systems~\cite{schedl2015music}. Studies on music listening behavior showed that personal traits and long-term music preferences are correlated as people tend to prefer music styles that align with their personalities~\cite{laplante2014improving,rentfrow2007content}. Furthermore, related work found a relationship between music and motivation~\cite{kim2020pepmusic}, music and emotion~\cite{juslin2001music,zentner2008emotions,juslin2004expression,yang2011music} or both personality and emotion~\cite{ferwerda2015personality}. 
Openness, a personality trait from the Five Factor Model~\cite{goldberg1993structure}, has also been shown to positively influence a user's preference for music recommendations~\cite{tintarev2013adapting}. \rev{Specifically, the authors of \cite{tintarev2013adapting} found that people tend to prefer recommendations from different kinds of music (i.e., openness) rather than varied within a specific kind of music (i.e., diversity).}
Others showed that familiarity has a positive influence on music preferences~\cite{schubert2007influence,pereira2011music} and that music preferences may change over time~\cite{moore2013taste}.
\rev{Another strand of research on modeling users' music preferences leverages content features, e.g., acoustic features.
It has been shown that the distribution of acoustic features of a user's preferred genre substantially influences the user's choice of music within other genres~\cite{barone2017acoustic}.
Also, acoustic features have been utilized to model users' preferences under different contextual conditions, in order to refine recommendation quality~\cite{gong2020contextual}.
Based on tracks' acoustic features, the authors of \cite{zangerle_ismir18} identified several types of music, and subsequently modeled each user by linearly combining the acoustic features of the music types. In contrast to these works, we focus on using acoustic features of music tracks for modeling and clustering beyond-mainstream music. Additionally, we link these beyond-mainstream music clusters to music genres and users in our Last.fm data sample.}

\paragraph{\rev{\textbf{Long-tail recommendations.}}}
\rev{Related research~\cite{celma:springer:2010,Oord2013:DCM} has found that individual music consumption is biased towards popular music and that usage data for less popular music is scarce. Due to the scarcity problem, items with no or few ratings (i.e., long-tail items) have little chance of being recommended~\cite{celma2008hits}. As a consequence, users that particularly favor items with few ratings or interactions are less likely to be recommended those items that they like~\cite{celma2009music}. That is problematic because many users, from time to time, prefer niche music~\cite{goel2010anatomy}. Therefore, such users are not well served as a result of their preference for less popular items. 
That has been attributed to \emph{popularity bias}, which corresponds to over-representation of popular items in the recommendation lists~\cite{ekstrand2018all,brynjolfsson2006niches,jannach2015recommenders}. Abdollahpouri et al.~\cite{abdollahpouri2019unfairness} studied popularity bias in a dataset of movies (i.e., the MovieLens 1M dataset~\cite{harper2015movielens}) from the user perspective. Their study showed that commonly used recommendation techniques tend to deliver worse recommendations to users who prefer less popular movies. In our work~\cite{kowald2020unfairness}, we found evidence for popularity bias in a Last.fm dataset and showed that traditional personalized recommendation algorithms such as collaborative filtering deliver worse recommendations for consumers of niche music. In the present work, we aim to gain a deeper understanding of the behavior and preferences of this beyond-mainstreaminess user group. Thus, in contrast to existing works in long-tail recommendations, we focus on the user rather than the item perspective.}

\paragraph{\textbf{\rev{Popularity bias in} music recommender systems.}} 
Music recommender systems~\cite{schedl2015music} \rev{are crucial tools in online streaming services such as Last.fm, Pandora, or Spotify. They help users find music that is tailored to their preferences. The basis of music recommender systems are user models derived from users' listening behavior, user properties such as personality (e.g.,~\cite{cheng2016music}), content features of music, or hybrid combinations of both, e.g.,~\cite{kaminskas_etal:recsys:2013,Donaldson2007hybridmusicrec,aggarwal2016ensemble,zangerle2018content}}.  
\rev{As discussed earlier, due to insufficient amounts of usage data for less popular items, many music recommendation algorithms do not provide useful recommendations for consumers of less popular and niche items.} 
As a remedy, in~\cite{lee2011my}, an approach is suggested that divides music consumers into experts and novices according to their long tail distribution in their playlists. These experts are then converted to nodes with bidirectional links connecting all the experts. These links are created to perform link analysis on the graph and to assign fine-grained weights to songs. The presented approach helps add music from the long-tail into the recommendation list.
In our previous research~\cite{lex2020modeling}, we use a framework~\cite{kowald2017tagrec} that employs insights from human memory theory to design a music recommendation algorithm that provides more accurate recommendations than collaborative filtering-based approaches for three groups of users, i.e., low-mainstream, medium-mainstream and high-mainstream users. While the awareness of popularity bias in music recommender systems increases (e.g.,~\cite{bauer2019allowing}), the characteristics of music consumers whose preferences lie beyond popular, mainstream music are still not well understood. \rev{In the present work, we shed light on the characteristics of such beyond-mainstream music consumers and relate them to openness and diversity patterns as well as recommendation quality. With this, we aim to provide useful insights for creating novel music recommendation models that mitigate popularity bias.} 

\section{Preliminaries}
\label{s:method}
We investigate the characteristics of beyond-mainstream music listeners in a dataset mined \rev{from} Last.fm, a popular music streaming platform. We characterize the tracks in our dataset with acoustic features. Besides, we compare the recommendation accuracy of beyond-mainstream music listeners with the one of mainstream music listeners to motivate our subsequent analysis of the characteristics of beyond-mainstream music listeners.

\subsection{Acoustic Music Features}
\label{sec:acoustic_features}
For our analyses, we characterize music tracks using acoustic features that describe the content of a given track. Following the lines of, e.g.,~\cite{ism16,6844510,mcvicar2011mining,zangerle_ismir18}, we rely on \rev{acoustic features provided by the Spotify API} as a compact characterization of tracks\footnote{\url{https://developer.spotify.com/web-api/get-several-audio-features/}}. The following eight features are extracted from the audio signal of a track:
 
\begin{description}
 \item[Danceability] captures how suitable a track is for dancing and is computed based ``on a combination of musical elements including tempo, rhythm stability, beat strength, and overall regularity''.
 \item[Energy] describes the perceived intensity and activity of a track and is based on the dynamic range, perceived loudness, timbre, onset rate, and general entropy of a track.
 \item[Speechiness] captures the presence of spoken words in a track. High speechiness values indicate a high degree of spoken words (e.g., an audiobook), whereas medium values indicate tracks with both music and speech (e.g., rap music). Low values represent typical music tracks.
 \item[Acousticness] measures the probability that the given track only contains acoustic instruments. 
 \item[Instrumentalness] quantifies the probability that a track contains no vocals, i.e., the track is instrumental. 
 \item[Tempo] measures the rate of the track's beat in beats per minute.
 \item[Valence] describes the ``emotional positiveness'' conveyed by a track (i.e., cheerful and euphoric tracks reach high valence values). 
 \item[Liveness] measures the probability that a track was performed live, i.e., whether an audience is present in the recording. 
 \end{description}

\subsection{Enriched Dataset of Music Listening Events}\label{sec:raw_dataset}
To study characteristics of beyond-mainstream users and their listening preferences, we create a novel dataset \rev{called} \emph{LFM-BeyMS} that contains the required information for such analyses. We base our work on a dataset gathered from the Last.fm music platform, which we considerably enrich with the music tracks' acoustic features (see Section~\ref{sec:acoustic_features})~\cite{zangerle:tismir20}. 
Additionally, we combine this data with mainstreaminess information of Last.fm users (see Section~\ref{subsec:identify_lowms}) as well as music genre information to identify beyond-mainstream listeners and music (see Section~\ref{s:beyond-music}).  

\begin{figure}[t!]
    \centering
    \includegraphics[width=0.8\textwidth]{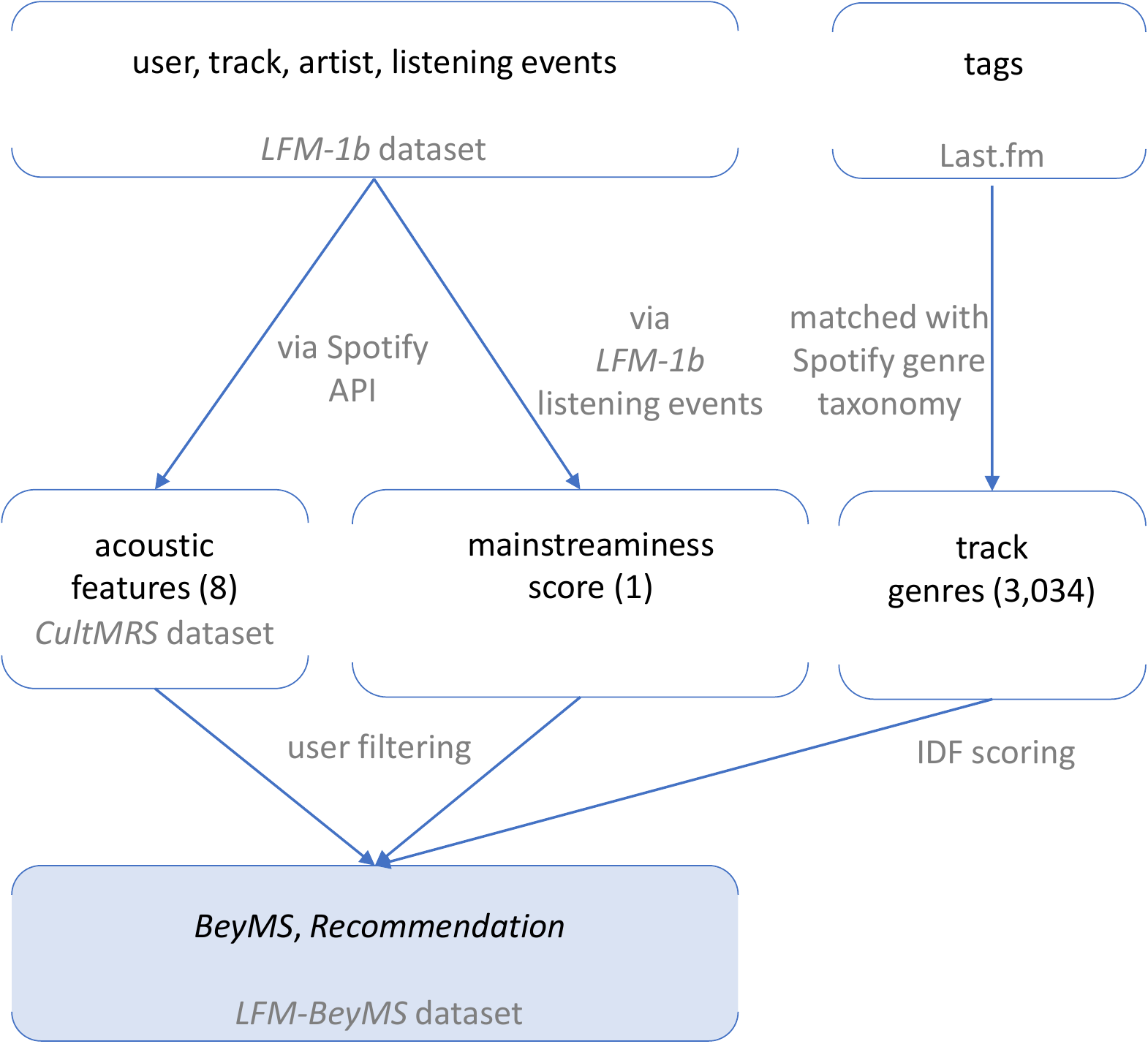}
    \caption{Overview of our new \emph{LFM-BeyMS} dataset and its data sources. We depict the different features, their origin, and relation, and show the feature groups with the number of contained features in brackets. \emph{LFM-BeyMS} contains \emph{BeyMS}, i.e., data to study the beyond-mainstream user group, and \emph{Recommendation}, i.e., data to conduct recommendation experiments of beyond-mainstream and mainstream music listeners.}
    \label{fig:dataset_overview}
\end{figure}

An overview of our new \emph{LFM-BeyMS} dataset and its data sources is depicted in Figure~\ref{fig:dataset_overview}. 
As shown, the starting point for our new dataset is the publicly available \emph{LFM-1b} dataset\footnote{\url{http://www.cp.jku.at/datasets/LFM-1b/}} of music listening information shared by users of the online music platform Last.fm~\cite{schedl2016lfm}. \emph{LFM-1b} contains listening histories of 120,322 users;
their listening records (or ``listening events'') have been created between January 2005 and August 2014. They sum up to over 1.1 billion listening events (LEs), where each LE is described by an (anonymous) user identifier, the artist name, the album name, the track name, and the timestamp of the listening event.
Also, the \emph{LFM-1b} dataset includes demographics of some users (i.e., country, age, and gender).

To enrich the \emph{LFM-1b} dataset to suit our task, we utilize our previously created \emph{CultMRS} music recommendation dataset~\cite{eva_zangerle_2019_3477842}. This dataset contains 55,191 users, who have listened to a total of 26,022,625 distinct tracks, captured by a total of 807,890,921 listening events~\cite{zangerle:tismir20}.

To further enrich the dataset with music acoustic features, we gather the acoustic features described in Section~\ref{sec:acoustic_features} for the tracks remaining in the dataset after the filtering described above. To this end, we rely on the Spotify API to gather content-based acoustic features for each track. Particularly, we search tracks using the $<$track, artist, album$>$ triples extracted from the \emph{LFM-1b} dataset using the Spotify search API\footnote{\url{https://developer.spotify.com/web-api/search-item/}} to gather the Spotify track URI of each track by using all three parts of the triple in a conjunctive query. In total, this allowed gathering 4,326,809 Spotify URIs. For the remainder of the tracks, we were not able to retrieve a URI. We attribute this to two factors: either the searched track is not provided by Spotify or the track, artist, and album information cannot be matched to a Spotify track unambiguously. Subsequently, we use the obtained track URI to query the acoustic features API, which returns the acoustic features of a given track (cf.~Section~\ref{sec:acoustic_features}). 
In a subsequent cleaning step, we remove all tracks for which the Spotify API did not provide the full set of acoustic features. 

That procedure provides us with a set of 3,478,399 unique tracks and their acoustic features. Within the LFM-1b dataset, this amounts to 13.36\% of the distinct tracks. 
Overall, these account for as much as 48.89\% of all listening events (i.e., the tracks listened to by users) of the LFM-1b dataset. The resulting dataset, now enriched by acoustic music descriptors, comprises a total of approximately 394 million listening events of 55,149 users. In Table~\ref{tab:raw_dataset} (column ``\emph{CultMRS}''), we provide further descriptive statistics of the \emph{CultMRS} dataset. We refine this dataset to create our new \emph{LFM-BeyMS} dataset (column ``\emph{BeyMS} in Table~\ref{tab:raw_dataset}), which consists of \emph{BeyMS}, i.e., data to study the characteristics of beyond-mainstream music listeners, and \emph{Recommendation}, i.e., data to conduct recommendation experiments of beyond-mainstream and mainstream music listeners. 

\begin{table}[t!]
\centering
\caption{Descriptive statistics of the \emph{CultMRS} dataset and our novel \emph{LFM-BeyMS} dataset. \emph{CultMRS} comprises acoustic features of tracks. \emph{LFM-BeyMS} is based on \emph{CultMRS} and consists of \emph{BeyMS} and \emph{Recommendation}. Our analyses of beyond-mainstream music listeners utilize \emph{BeyMS} and our recommendation experiments utilize~\emph{Recommendation}, which includes listening events of both users with beyond-mainstream and mainstream music taste.}
\resizebox{\textwidth}{!}{
\begin{tabular}{l| r r r}
\toprule
\multirow{2}{*}{Item} & \multirow{2}{*}{\emph{CultMRS}~\cite{eva_zangerle_2019_3477842}} & \multicolumn{2}{c}{\emph{LFM-BeyMS} (our novel dataset)}\\
\cmidrule{3-4}
 &  & \emph{BeyMS} & \emph{Recommendation}\\
\midrule
Users & 55,149 & 2,074 & 4,148 \\
Tracks & 3,478,399 & 157,444 & 1,084,922 \\
Artists & 337,840 & 14,922 & 110,898 \\
Listening Events (LEs) & 394,944,868 & 4,916,174 & 16,687,363 \\
Min. LEs per user & 1 & 3 & 9 \\
$Q_1$ LEs per user & 1,442 & 1,254 & 2,604 \\
Median LEs per user & 5,667 & 2,048 & 3,766 \\
$Q_3$ LEs per user & 9,738 & 3,239 & 5,252 \\
Max. LEs per user & 399,210 & 10,536 & 11,177 \\
Avg. LEs per user & 7,161.41 ($\pm$ 10,326.91) & 2,371.526 ($\pm$ 1,520.629) & 4,022.990 ($\pm$ 1,898.060) \\
\bottomrule
\end{tabular}}
\label{tab:raw_dataset}
\end{table}

%%%%%%%%%%%%%%%%%%%%%%%%%%%%%%%%%%%%%%%%%%%%%%%%%%%%%%%%%%%%%%%%%
\subsection{Identifying Beyond-Mainstream Music Listeners}
\label{subsec:identify_lowms}
To identify beyond-mainstream music listeners, for each user, we compute a mainstreaminess score, which is generally defined as the overlap between a user's individual listening history and the aggregated listening history of all Last.fm users in the dataset. In this vein, the mainstreaminess score reflects a user's inclination to music listened to by the Last.fm mainstream listeners (i.e., the ``average'' Last.fm listener in the dataset). 
In \cite{10.1371/journal.pone.0217389}, several measures of user mainstreaminess are defined. 
Out of these, we choose the \emph{M-global-R-APC} definition since it yielded good results in context-based music recommendation experiments for the \emph{LFM-1b} dataset, as evidenced in \cite{10.1371/journal.pone.0217389}. 
The \emph{M-global-R-APC} measure approximates a user's mainstreaminess score by computing Kendall's $\tau$~\cite{kendall1938new} rank correlation between the user's vector of artist play counts and the global vector of artist play counts (aggregated over all users in the dataset). This definition also explains the name of the measure, where ``M'' refers to mainstreaminess, ``global'' indicates the global perspective, ``R'' stands for rank correlation, and ``APC'' refers to artist play counts.

Next, we describe how we identify our beyond-mainstream users via filtering the users by the number of listening events (see Figure~\ref{fig:le_dist} and Section~\ref{s:distribution_le}) and by mainstreaminess scores (see Figure~\ref{fig:mainstreaminess_dist} and Section~\ref{s:distribution_mainstreaminess}). 

\subsubsection{Filtering Users by the Number of Listening Events} \label{s:distribution_le}
For our study, we select the users so that listeners of 
different levels of ``listening activity'' are equally represented. 
We conduct a Gaussian kernel density estimation (KDE)~\cite{sheather2004density} on the distribution of listening events over users to estimate the continuous probability density function (PDF)~\cite{davis2011remarks}.
However, KDE estimates the PDF via discrete bins and hence, it is necessary to approximate the gradient via the principle of finite differences.
The gradient of the PDF helps us identifying regions of increasing or decreasing probability.

\begin{figure}[t!]
    \includegraphics[width=0.99\textwidth]{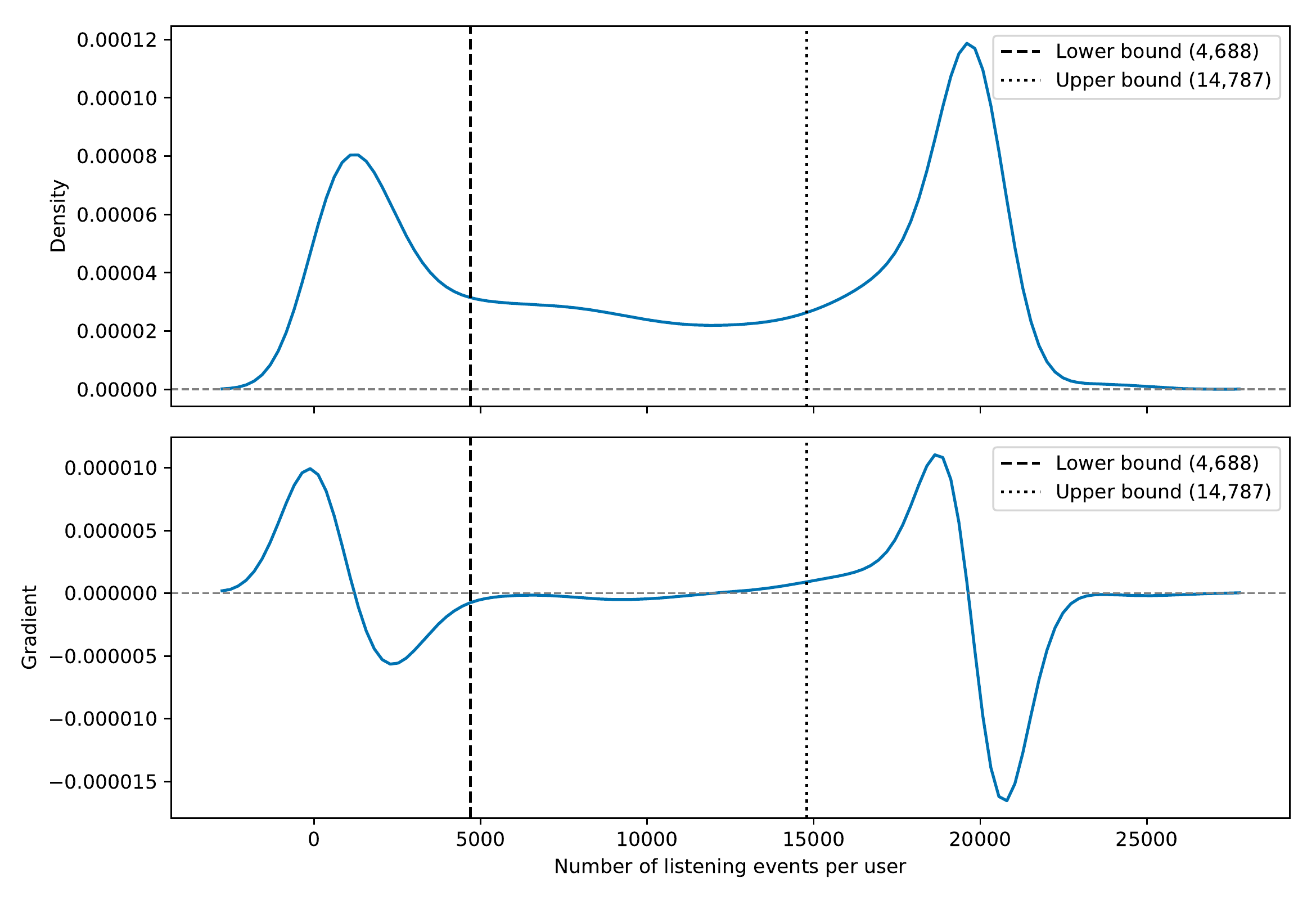}
    \caption{Distribution of listening events in our set of Last.fm users. We set the lower and upper bound marked as dashed and dotted lines, respectively based on the gradient, which results in 12,814 users with a similar number of listening events.}
    \label{fig:le_dist}
\end{figure}

Figure~\ref{fig:le_dist} shows that two large subsets of users exist that exhibit either very few or an abundance of listening events.
For our analysis, we consider only users who are not in one of the subsets as mentioned \rev{earlier. On the one hand}, we exclude users with too little data available for studying their listening behavior; and on the other hand, we exclude so-called power listeners who might bias our analyses. Furthermore, such users with a very high number of listening events are often radio stations, which do not contribute reliable data to our investigations.

Hence, we define lower and upper bounds regarding the number of users' listening events to include in our study, such that the rate of change in terms of the number of listening events is minimal and stable within these boundaries.
That requires the gradient of the region within the lower and upper bound to be near zero (i.e., $\pm 10^{-6}$).
By computing the second-order accurate central differences~\cite{quarteroni2007numerical}, we obtain an approximation of the gradient and find the longest cohesive region fulfilling the requirements between a lower bound of 4,688 and an upper bound of 14,787 listening events per user, which leads to 12,814 users.

\subsubsection{Filtering Users by Mainstreaminess Scores}
\label{s:distribution_mainstreaminess}
\begin{figure}[t!]
    \centering
    \includegraphics[width=1.0\textwidth]{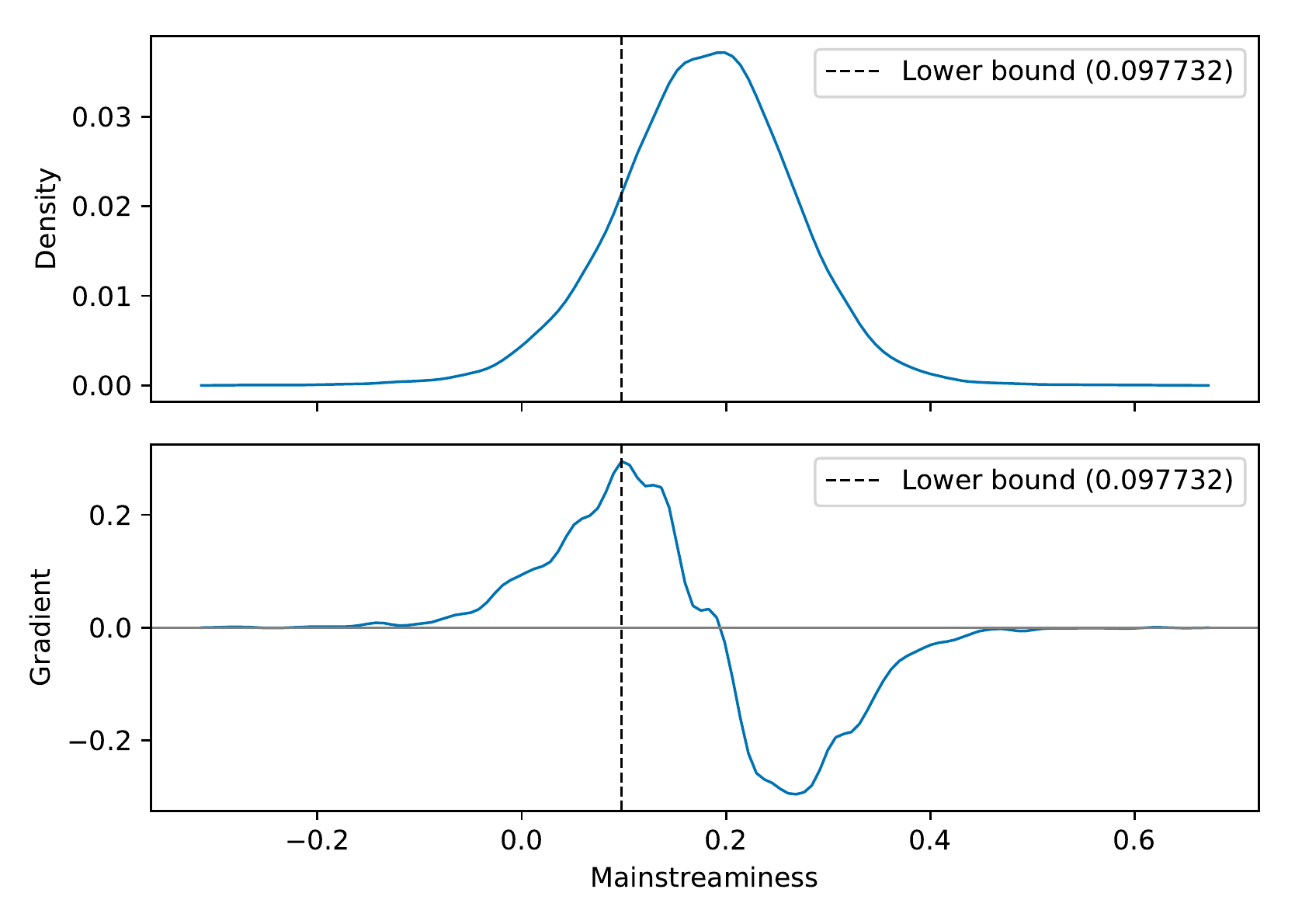}
    \caption{Mainstreaminess distribution of the 12,814 users illustrated in Figure~\ref{fig:le_dist}. Based on the maximum gradient, we select an upper bound of 0.097732 to identify the 2,074 beyond-mainstream users of the \emph{BeyMS} user group.}
    \label{fig:mainstreaminess_dist}
\end{figure}

Figure~\ref{fig:mainstreaminess_dist} illustrates the mainstreaminess distribution of the 12,814 users that we have extracted based on the number of listening events. Here,  mainstreaminess is defined according to the \emph{M-global-R-APC} definition taken from~\cite{10.1371/journal.pone.0217389} (explained in Section~\ref{subsec:identify_lowms}). 
By setting an appropriate upper bound, we aim to exclude mainstream music listeners. 
In other words, we aim to set the upper bound to the beginning of the distribution's bulk, which is motivated as follows:
Firstly, the first inflection point (i.e., maximal gradient) of a Gaussian distribution is found at $\mathbb{E}[X] - std(X)$, where 
$\mathbb{E}[X]$ is the expectation, and $std(X)$ is the standard deviation of the Gaussian random variable~$X$.
Secondly, the first inflection point of a Gaussian distribution is equivalent to the 15.9-percentile.
By setting the mainstreaminess threshold to this point, we intend to omit the majority of users and hence, only consider the 15.9\% of users with the lowest mainstreaminess scores.
Utilizing this upper bound on the mainstreaminess score, we obtain a set of 2,074 beyond-mainstream users. 
Furthermore, the Gaussian assumption can be strengthened by the observation that the 2,074 beyond-mainstream users represent 16.19\% of users. In the remainder of this paper, we refer to this set of beyond-mainstream music listeners as \emph{BeyMS}.

%%%%%%%%%%%%%%%%%%%%%%%%%%%%%%%%%%%%%%%%%%%%%%%%%%%%%
\subsection{Identifying Beyond-Mainstream Music} \label{s:beyond-music}
We aim to study beyond-mainstream listeners in terms of their music taste. We characterize music via its acoustic features, as described in Section~\ref{sec:acoustic_features}, and also investigate genres as an alternative way to describe a music track via conventional categories. As the \emph{LFM-1b} dataset does not contain genre annotations of tracks \rev{and the Spotify API only provides genres on artist level\footnote{\url{https://developer.spotify.com/documentation/web-api/reference-beta/\#endpoint-get-an-artist}}}, we leverage the tags assigned to each track by Last.fm users to identify genre annotations. To obtain these tags, we use the respective Last.fm API endpoint\footnote{\url{https://www.last.fm/api/show/track.getTopTags}}. 
After having fetched the tags for each track, we de-capitalize them and remove all non-alpha-numeric characters.
Since not all tags used by Last.fm users correspond to actual music genres 
(e.g., the ``seenlive'' tag is used to indicate that a user has seen an artist performing this track live), we use a fine-grained music genre taxonomy consisting of 3,034 genres that are also utilized by Spotify, which we gather from the EveryNoise service (2019-07-24)\footnote{\url{http://everynoise.com/}}. 
Specifically, for each track listened to by any of our \emph{BeyMS} users, we remove all tags that are not part of the EveryNoise genre taxonomy, using a case-insensitive matching approach.

\begin{figure}[t!]
    \centering
    \includegraphics[width=0.8\textwidth]{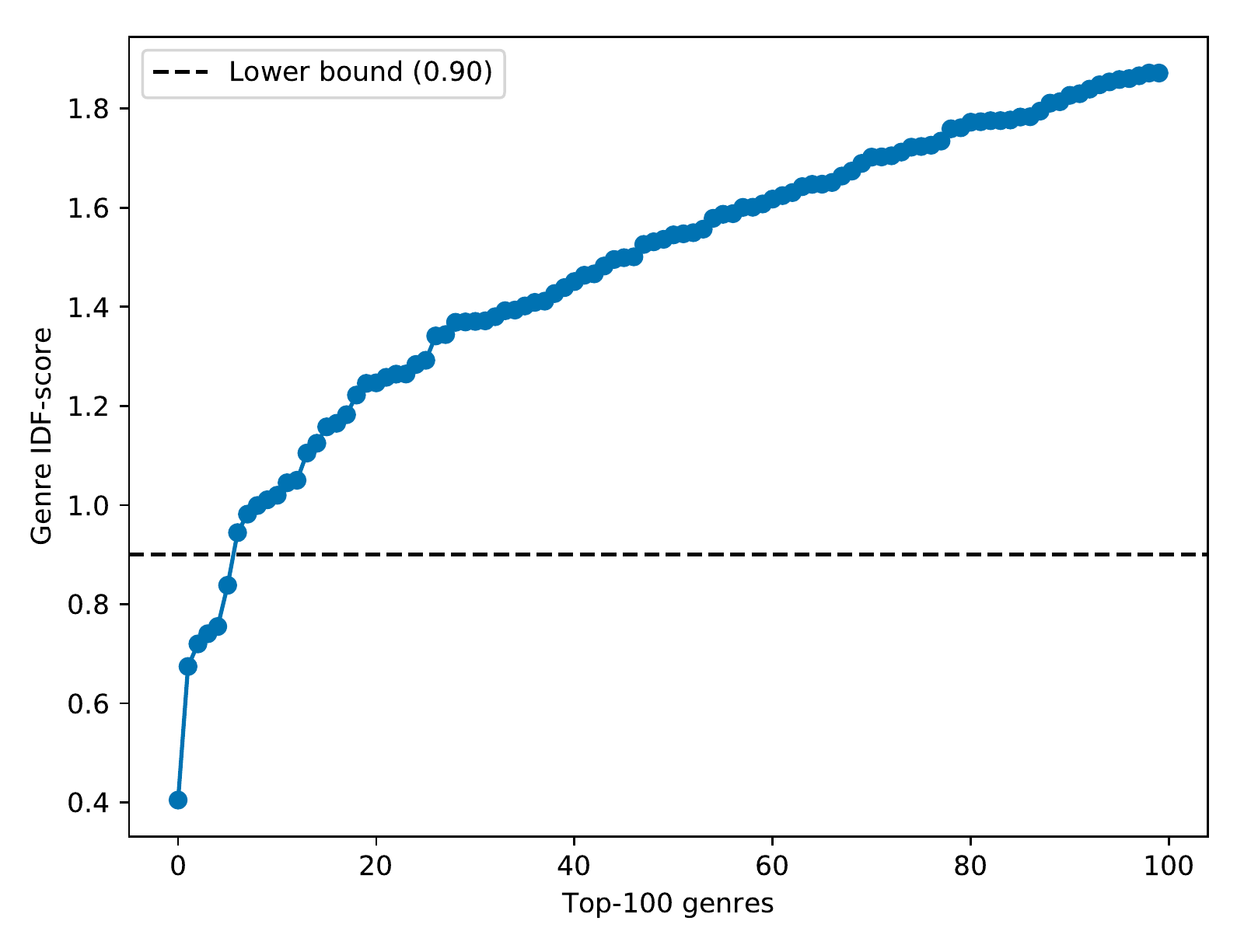}
    \caption{IDF-score distribution of the top-100 genres in ascending order (i.e., from coarse-grained to fine-grained). The 6 coarse-grained genres below the lower bound of 0.90 are removed from the genre assignments, i.e., ``rock'', ``pop'', ``electronic'', ``metal'', ``alternativerock'', ``indierock''.}
    \label{fig:genre_IDF}
\end{figure}

We note that Last.fm users tend to assign very general genre tags to a large number of tracks, such as ``pop'' or ``rock''. To remove these coarse-grained genres and to identify fine-grained beyond-mainstream music genres, we calculate the inverse document frequency (IDF)~\cite{jones1972statistical} metric of our genre-track distribution by treating genres as terms and tracks as documents, i.e.,  $IDF(g) = \log_{10}\frac{|T|}{|\{t \in T \mathrm{\,with\,} g \in G_t\}|}$. 
More precisely, the IDF-score of genre $g$ is determined by relating the number of all tracks $|T|$ to the number of tracks annotated with genre $g$ where $|G_t|$ is the set of genres assigned to track $t$. This way, a coarse-grained genre receives a small IDF-score, while a fine-grained genre receives a high IDF-score.
Figure~\ref{fig:genre_IDF} shows the IDF-score distribution of the top-100 genres in ascending order (i.e., from coarse-grained to fine-grained).
Here, we identify two groups of genres, where the first group consists of 6 genres with small IDF-scores, and the second group consists of 94 genres with high IDF-scores.
The visual inspection of Figure~\ref{fig:genre_IDF} indicates that the lower bound of 0.90 serves as a discriminant between these two groups of coarse-grained and fine-grained genres.
Consequently, we remove the 6 coarse-grained genres (i.e., ``rock'', ``pop'', ``electronic'', ``metal'', ``alternativerock'', ``indierock'') from the genre assignments of our tracks, which leads to 157,444 \rev{out of 799,659 tracks listened to by \emph{BeyMS} users} with at least one remaining genre. In total, these tracks are annotated with 1,418 unique genre identifiers. 

\rev{We are aware of the fact to our track filtering procedure leads to incomplete listening profiles of users. Since we rely on genres to describe beyond-mainstream music, these filtering steps are necessary for our study. To ensure that the \emph{BeyMS} users' reduced listening profiles are still representative of their music preferences, we further investigate the consequences of the filtering procedure. 
Here, we find that a user's listening history (i.e., the entirety of a user's listening events) is reduced to 61\% on average. 
However, we also find that there are only 62 of the 2,074 \emph{BeyMS} users, for whom the listening history is reduced to less than 20\%. For these users most affected by the filtering, we compare the acoustic feature distributions of their listened tracks before and after the filtering steps, and find that filtering only marginally affects the acoustic feature distributions (i.e., average change in mean $= 0.0098 \pm 0.0148$). This means that the acoustic feature distribution contained in the user's profile is highly robust against the filtering.} 
The statistics of \emph{BeyMS} are summarized in column ``\emph{BeyMS}'' in Table~\ref{tab:raw_dataset}. 

\subsection{Recommendations for Beyond-Mainstream Music Listeners} \label{s:rec_first}
In order to compare the recommendation accuracy of recommendations received by the users of our \emph{BeyMS} group 
and by mainstream users, we construct a dataset consisting of \emph{BeyMS}'s listening events and the listening events of an equally-sized group of mainstream users.
Therefore, we define the \emph{MS} user group as 2,074 (i.e., the size of our \emph{BeyMS} group) randomly-chosen users with a mainstreaminess score that is higher than the upper bound for low mainstreaminess, identified in Figure~\ref{fig:mainstreaminess_dist}. 
Furthermore, the \emph{MS} users are also in between the lower and upper bounds for listening events identified in Figure~\ref{fig:le_dist}. 
As shown in Table~\ref{tab:raw_dataset} (column ``\emph{Recommendation}''), the dataset used for the evaluation of recommendations 
contains data of 4,148 distinct \emph{BeyMS} and \emph{MS} users, 1,084,922 distinct tracks, and 16,687,363 listening events. 

We use the Python-based open-source recommendation library Surprise\footnote{\url{http://surpriselib.com/}} to compute and evaluate recommendations. \rev{One advantage of using Surprise is that it provides built-in recommendation algorithms as well as a standardized evaluation pipeline, which enhances the reproducibility of our research.} Since Surprise is focused on rating prediction, we formulate our music recommendation scenario also as a rating prediction problem, in which we predict the preference of a target user $u$ for a target track $t$. As done in~\cite{schedl2017distance}, we model the preference of $t$ for $u$ by scaling the play count (i.e., number of listening events) of $t$ by $u$ to a range of [1; 1,000] using min-max normalization. \rev{We perform this normalization on the individual user level to ensure that all users share the same preference value ranges}.  
Thus, with this method, we ensure that each user's most listened track has a preference value of 1,000, while their least listened track has a preference value of 1. 
\rev{To ensure that this min-max normalization procedure does not disrupt the play count distribution of our users, we compare the original play count distribution with the normalized distribution and find that both distributions are strongly right-skewed. Specifically, we find very similar distributions for large amounts of our play count data.}

We utilize a selection of \rev{Suprise's built-in} recommendation methods consisting of one baseline approach (i.e., UserItemAvg), two neighborhood-based approaches (i.e., UserKNN and UserKNNAvg), and one matrix factorization-based approach (i.e., NMF). 
Specifically, UserItemAvg predicts the average play count in the dataset by also accounting for deviations of $u$ and $t$, for example, if a user~$u$ tends to have more listening events than the average Last.fm user~\cite{koren2010factor}. UserKNN~\cite{herlocker2004evaluating} is a user-based collaborative filtering approach and is calculated using $k$~=~40 
nearest neighbors and the cosine similarity metric, which are the default settings of Surprise. UserKNNAvg is an extension of UserKNN~\cite{herlocker2004evaluating} that also takes the average rating of target user $u$ into account.
Finally, NMF, i.e., non-negative matrix factorization~\cite{luo2014efficient}, is calculated using 15 latent factors, which is the default parameter in the Surprise library. 
\rev{As shown in our previous work~\cite{kowald2020unfairness}, NMF is also capable of recommending non-popular items from the long tail and should therefore especially be of interest for our beyond-mainstream recommendation setting.}

We use Surprise's default parameters and refrain from performing any hyperparameter tuning since we are only interested in assessing (relative) performance differences between the two user groups \emph{BeyMS} and \emph{MS}, and not in outperforming any state-of-the-art algorithm. This is also the reason why we focus on traditional algorithms instead of investigating the most recent deep learning architectures, which would also require a much higher computational effort. 

\begin{table}[t!]
    \centering
    \caption{Mean absolute error (MAE) results for the two user groups \emph{MS} and \emph{BeyMS} of different mainstreaminess and a selection of standard recommendation algorithms. A one-tailed Mann-Whitney-U test ($\alpha = .0001$) provides significant evidence, indicated by ***, that all algorithms perform worse on \emph{BeyMS} than on \emph{MS} in terms of MAE. Furthermore, NMF (as shown in bold) outperforms the other three approaches UserItemAvg, UserKNN and UserKNNAvg.} 
    \begin{tabular}{l | l l l l}
        \toprule
        User group & UserItemAvg & UserKNN & UserKNNAvg & NMF \\ \midrule
        \emph{BeyMS} & 63.4608*** & 71.6694*** & 67.5770*** & \textbf{57.7703}*** \\
        \emph{MS} & 61.2562 & 68.4894 & 63.3985 & \textbf{54.8182} \\ \midrule
        Overall & 62.2315 & 69.8962 & 65.2469 & \textbf{56.2492} \\ \bottomrule
    \end{tabular}
    \label{tab:mae_mainstr_group}
\end{table}

The resulting mean absolute error (MAE) results can be observed in Table~\ref{tab:mae_mainstr_group} (and correspond to the ones already shown in Figure~\ref{fig:motivation_example}). 
We favor MAE over the commonly used root mean squared error (RMSE) due to several pitfalls, especially regarding the comparison of groups with different numbers of observations~\cite{willmott2005advantages}.
Here, we perform 5-fold cross-validation leading to 5 different 80/20 train-test splits and average the MAE over the 5 folds. 
NMF clearly outperforms UserItemAvg as well as the two neighborhood-based methods (i.e., UserKNN and UserKNNAvg) both for the two user groups (see rows ``\emph{BeyMS}'' and ``\emph{MS}'') separately and overall without distinguishing between the user groups (see row ``Overall''). 
Additionally, we conduct a one-tailed Mann-Whitney-U test ($\alpha = .0001$), where we define the null-hypothesis as the MAE for \emph{MS} being larger than or equal to the MAE for \emph{BeyMS}. 
Results marked with *** indicate that the null-hypothesis was rejected for every fold.
Thus, all algorithms \rev{(including NMF)} provide a significantly larger error for \emph{BeyMS} than for \emph{MS}.
In other words, recommendation quality is significantly better for users with  mainstream taste than for users who prefer beyond-mainstream music across all recommendation approaches.

These initial results underpin the need to study the characteristics of the \emph{BeyMS} user group that receives worse recommendations. The corresponding experiments are presented in the next section. 

\section{Characteristics of Beyond-Mainstream Music \rev{and} Listeners}
\label{s:results}
\rev{We identify the types of beyond-mainstream music using unsupervised clustering and characterize these types with respect to acoustic features and music genres. Besides, we detect subgroups of beyond-mainstream music listeners by assigning users to these clusters and evaluate the recommendation quality obtained for these subgroups. Finally, we discuss the recommendation quality with respect to openness and diversity.} For this, we relate to the definitions given by~\cite{tintarev2013adapting}:
\begin{description}
\item[Openness] is the across-groups diversity (or categorical diversity) and describes if users of one group also listen to the music of other groups.
\item[Diversity] is the within-groups diversity (or thematic diversity) and describes the dissimilarity of music listened to by users within groups.
\end{description}
Based on the findings of~\cite{tintarev2013adapting}, we would expect that subgroups with high openness should receive more accurate recommendations than subgroups with high diversity. 

\subsection{Clustering \rev{and Characterizing} Beyond-Mainstream Music} \label{s:clustering}
To study the different types of music listened to by the users in our \emph{BeyMS} group, we conduct a cluster analysis. 
Specifically, we cluster the 157,444 tracks listened to by \emph{BeyMS} users, where each track is described by the eight acoustic features danceability, energy, speechiness, acousticness, instrumentalness, tempo, valence, and liveness (see Section~\ref{sec:acoustic_features}). We scale the value ranges of these features to [0, 1] using min-max normalization. 
The use of latent representations of musical elements such as tracks was shown to be efficient in the area of music information retrieval~\cite{moore2012learning,levy2008learning,zangerle_ismir18}.
Furthermore, for visually analyzing the obtained music clusters and decreasing computation time, we favor a reduction of dimensionality to two dimensions.

We conduct experiments with a broad body of dimensionality reduction methods, i.e., linear and nonlinear 
principal component analysis (PCA)~\cite{tipping1999mixtures}, locally linear embedding~\cite{roweis2000nonlinear}, multidimensional scaling~\cite{kruskal1964nonmetric}, Isomap~\cite{tenenbaum2000global}, spectral embedding~\cite{ng2002spectral}, t-distributed stochastic neighbor embedding (t-SNE)~\cite{maaten2008visualizing} and uniform manifold approximation and projection (UMAP)~\cite{mcinnes2018umap}. 
We visually inspected the 2-dimensional feature spaces created by these methods with regards to the clustering quality, and we obtained the visually most homogeneous results with UMAP. 
Moreover, UMAP has already been successfully used in the music domain~\cite{zangerle_ismir18} and thus, we use it for the remainder of our experiments. 
Specifically, we utilize the open-source implementation of UMAP~\cite{mcinnes2018umap-software}, which requires four parameters: (i)~the distance metric $M$ in the input space, (ii)~the number of latent dimensions $D$, (iii)~the minimum distance of points in the latent space $d_{min}$, and (iv)~the number of neighbors of a point~$N$. Based on experimentation and related literature (e.g.,~\cite{mcinnes2018umap-software}), we set the distance metric $M$ to the Euclidean distance, the number of latent dimensions $D$ to 2, the distance $d_{min}$ to 0.1 and the number of neighbors $N$ to 15.

\begin{figure}[t!]
    \centering
    \includegraphics[width=0.8\textwidth]{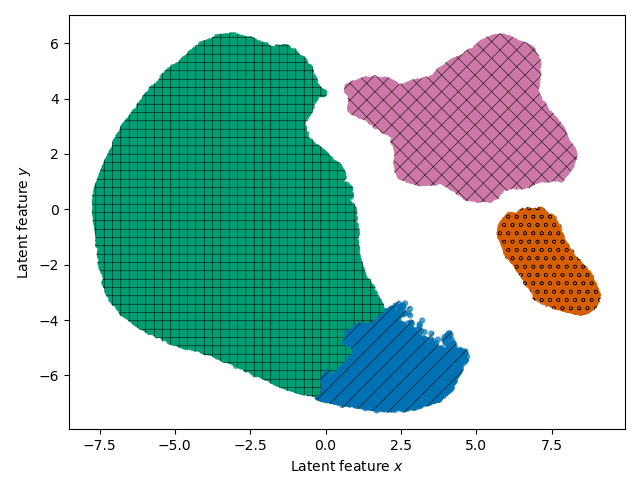}
    \caption{Music clustering results obtained with  HDBSCAN* and UMAP for the 2-dimensional mapping. The outputs are four clusters with the following cluster sizes: 12,148 (\rev{blue, hatch: /}), 92,798 (\rev{green, hatch: +}), 7,629 (\rev{orange, hatch: o}) and 30,379 (\rev{pink, hatch: x}) tracks. 14,490 of our 157,444 \emph{BeyMS} tracks have not been assigned to a cluster.}
    \label{fig:track_cluster}
\end{figure}

In a next step, we perform clustering on the dimensionality-reduced acoustic features of tracks. 
Again, we conduct experiments with various clustering methods, i.e., DBSCAN~\cite{ester1996density}, $K$-Means~\cite{bishop2006pattern}, Gaussian mixture models~\cite{reynolds2015gaussian}, affinity propagation~\cite{frey2007clustering}, spectral clustering~\cite{shi2000normalized}, hierarchical agglomerative clustering~\cite{murtagh2014ward}, OPTICS~\cite{ankerst1999optics} and HDBSCAN*~\cite{mcinnes2017accelerated}.
Here, we obtain the best results with respect to cluster cohesion and separation using HDBSCAN*. 
Furthermore, HDBSCAN* was also already used by related work to cluster music items~\cite{yoo2017data}.
We employ the open-source implementation of HDBSCAN*~\cite{McInnes2017} that requires four parameters: (i)~the minimum cluster size $s_{min}$ that defines the minimum size of a group of points to consider a cluster, (ii)~the minimum number of samples in the neighborhood of a core point $N_{min}$, which quantifies how conservative the clustering is, (iii)~$\varepsilon$, which enables the recovery of DBSCAN clusters if the $s_{min}$ value is not reached, and (iv)~the scaling of the distance $\alpha$, which is another measure of the clustering's conservativeness. 
In detail, $\alpha$ scales the distance between two points, which determines whether these points are merged into a cluster. 
This scaling is used in the construction of HDBSCAN*'s hierarchy of clusterings. 
Again, we find the best-suited parameters based on experimentation and related literature (e.g.,~\cite{mcinnes2017accelerated}). 
Specifically, we require each cluster to comprise a sufficiently large number of tracks to increase the level of significance of our subsequent experiments.
We expect the existence of very small music clusters and thus, search for the optimal value of the minimal cluster size $s_{min}$ in the search space of $\{1,\!000; 1,\!025; \dots; 1,\!475; 1,\!500\}$,  where we obtain the best results with respect to the within-cluster variance for $s_{min} = 1,375$. 
Furthermore, tightly packed clusters without any contribution of noise should be favored. 
In other words, all points within a cluster should be within the neighborhood of at least one core point.
Thus, we set the minimal number of samples in the neighborhood $N_{min} = s_{min} = 1,375$.
The remaining two parameters are set to their default values, i.e., $\varepsilon$ = 0 and $\alpha$ = 1. 

Figure~\ref{fig:track_cluster} shows the results of the clustering process using HDBSCAN* and UMAP for the 2-dimensional mapping. 
This process leads to four music clusters.
Here, \rev{the green cluster (hatch: +)} is the largest one with 92,798 tracks, followed by \rev{the pink cluster (hatch: x)} with 30,379 tracks and \rev{the blue cluster (hatch: /)} with 12,148 tracks. 
The smallest cluster is \rev{the orange one (hatch: o)} as it contains 7,629 tracks. The remaining 14,490 of our 157,444 \emph{BeyMS} tracks have not been assigned to a cluster and thus, will not be included in further analyses and interpretations.
\rev{Next, we describe how we name these clusters based on their music genre distributions.}

%%%%%%%%%%%%%%%%%%%%%%%%%%%%%%%%%%%%%%%%%%%%%%%%%%%%
\begin{figure}[t!]
    \centering
    \includegraphics[width=\textwidth]{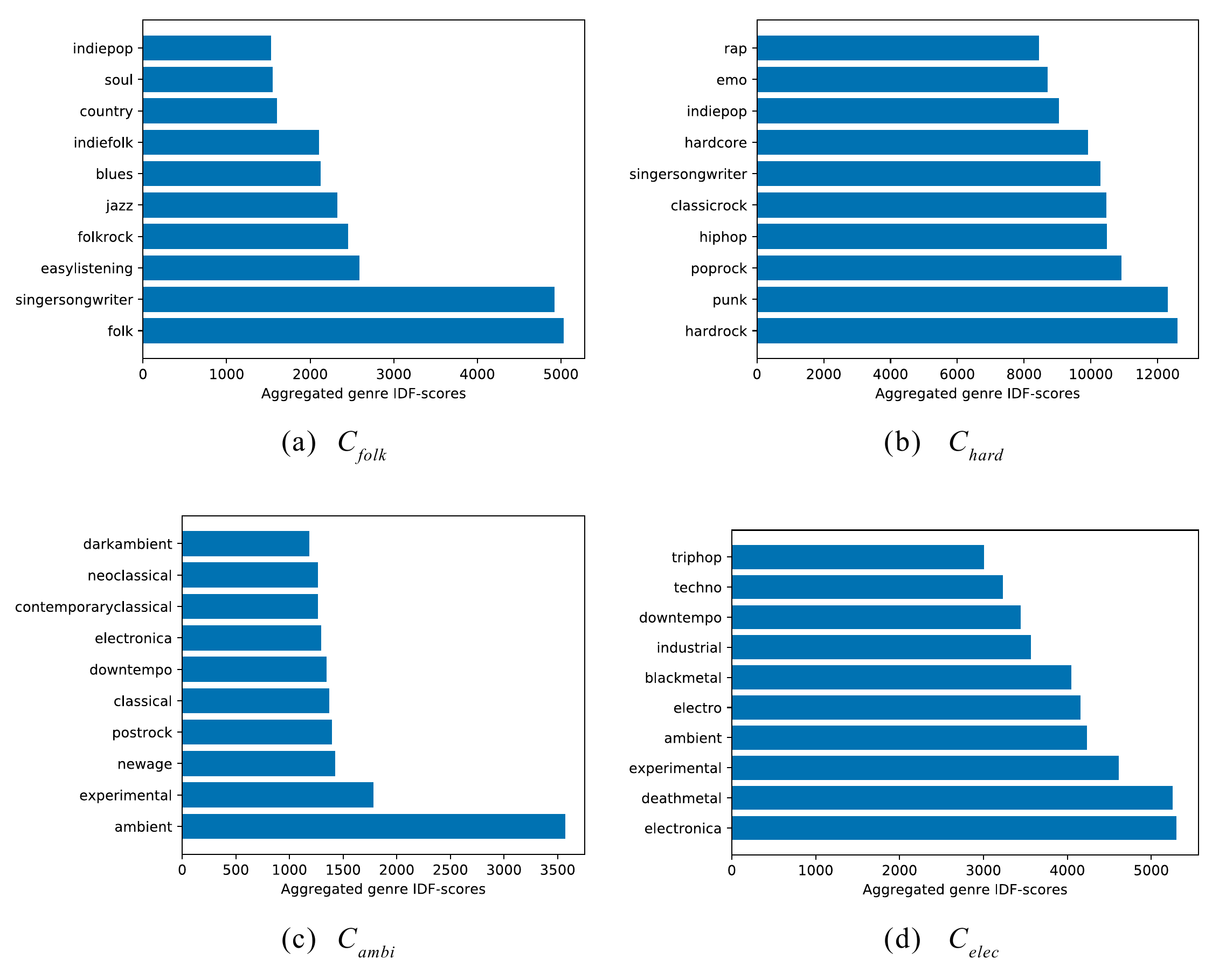}
    \caption{Top-$10$ genres of the four music clusters $C_1$--$C_4$ according to the aggregated genre IDF-scores. We name the clusters according to the top genre, i.e., (a) \rev{blue (hatch: /)} $\rightarrow$ $C_{folk}$ (``folk''), (b) \rev{green (hatch: +)} $\rightarrow$ $C_{hard}$ (``hardrock''), (c) \rev{orange (hatch: o)} $\rightarrow$ $C_{ambi}$ (``ambient''), and (d) \rev{pink (hatch: x)} $\rightarrow$ $C_{elec}$ (``electronica'').}
    \label{fig:tfidf_dists}
\end{figure}

\subsubsection{Genre Distributions}
In Figure~\ref{fig:tfidf_dists}, we illustrate the top-$10$ genres of the four music clusters. For this, we refer to the genre IDF-scores presented in Section~\ref{s:beyond-music} and weight each genre assigned to a track in a cluster with its corresponding IDF-score. For example, if a genre with an IDF-score of 1.4 is assigned to 1,000 tracks in a cluster, it is visualized as an aggregated genre IDF-score of 1,400 in the corresponding plot of Figure~\ref{fig:tfidf_dists}. Based on the genre distributions, we label each cluster according to its top genre.

With respect to \rev{the blue cluster (hatch: /)} in Plot (a), we find top genres such as ``folk'' and ``singersongwriter'', which 
typically reflect music with high acousticness. In the remainder of this paper, therefore, we refer to this cluster as $C_{folk}$. The top genres of \rev{the green cluster (hatch: +)} in Plot (b) are 
typical high energy music genres such as ``hardrock'', ``punk'', ``poprock'', and ``hiphop''. Based on this, we name this cluster $C_{hard}$.

For \rev{the orange cluster (hatch: o)} in Plot (c), we find genres that reflect music with high acousticness and high instrumentalness such as ``ambient'', ``experimental'', ``newage'', and ``postrock''. As ``ambient'' clearly dominates the genre distribution for this cluster, we name this cluster $C_{ambi}$. Similarly to $C_{folk}$, this cluster contains music with high acousticness; yet, while $C_{folk}$ is characterized by low instrumentalness music, $C_{ambi}$ is characterized by a high level of instrumentalness. 
Finally, Plot (d) shows the genre distribution of \rev{the pink cluster (hatch: x)} with ``electronica'' as the top genre, which leads to the name $C_{elec}$ for this cluster.

Thus, both, $C_{elec}$ and $C_{hard}$, consist of high energy music but in contrast to $C_{hard}$, $C_{elec}$ also comprise high instrumentalness values. This also makes sense when looking at other top genres of $C_{elec}$ such as ``deathmetal'' and ``blackmetal'' where guttural vocal techniques are often mistakenly classified as another type of instrument~\cite{york2004voices}.

\begin{figure}[t!]
    \centering
    \includegraphics[width=0.8\textwidth]{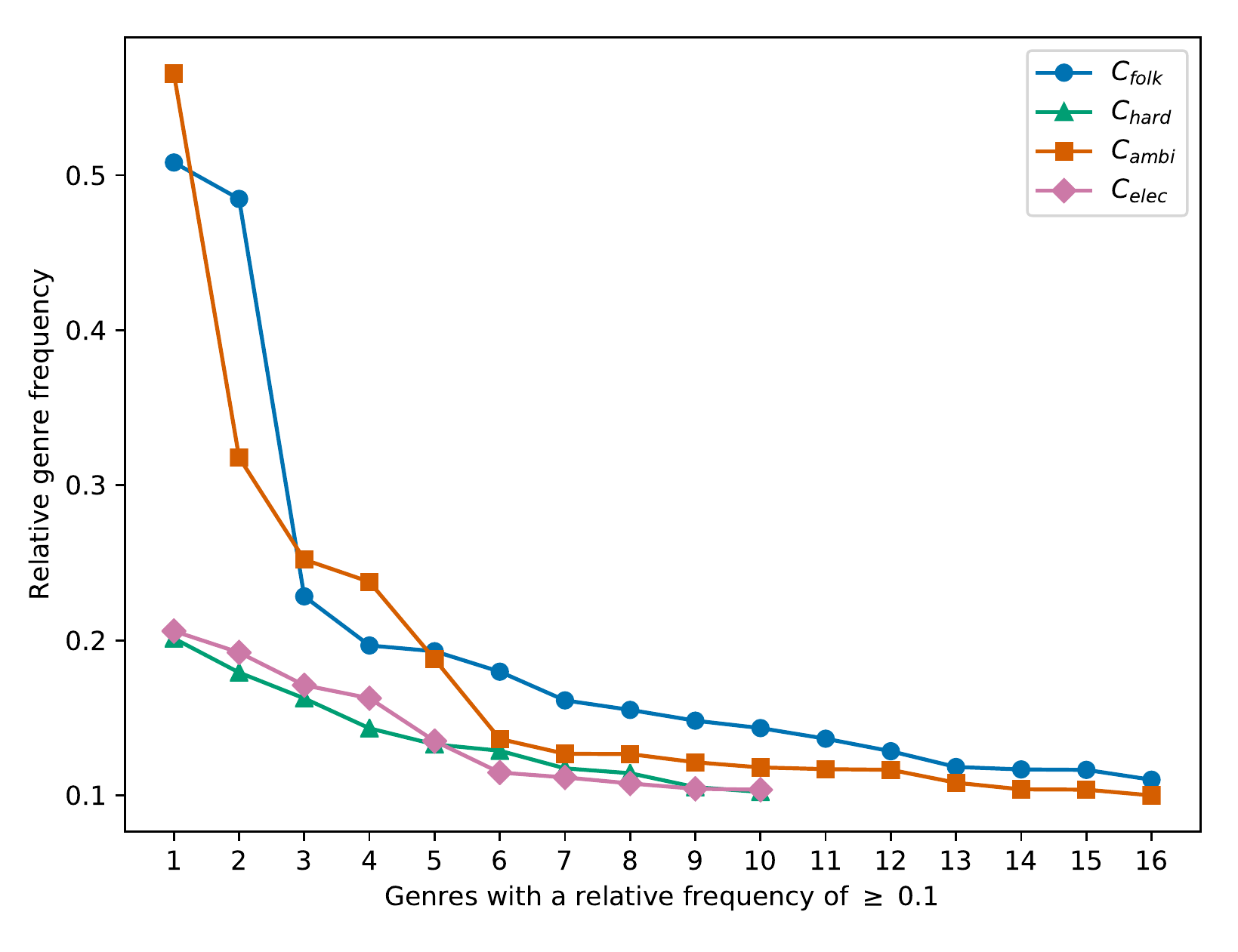}
    \caption{Relative genre frequency distribution of the four music clusters. While there are dominating genres in $C_{folk}$ and $C_{ambi}$, the genre distribution is more diverse in $C_{hard}$ and $C_{elec}$.}
    \label{fig:top10genres}
\end{figure}

To compare the genre distributions among the four music clusters, we illustrate the relative genre frequency distribution of the clusters in Figure~\ref{fig:top10genres}. 
The relative frequency of a genre $g$ depicts the fraction of listening events of tracks within a cluster $c$ that are annotated with $g$. Here, we only show genres with a minimum relative genre frequency of 0.1.
We see that there are clearly dominating genres in $C_{folk}$ and $C_{ambi}$, whereas the genre distributions in $C_{hard}$ and $C_{elec}$ are more evenly distributed. When relating this finding to the findings of Figure~\ref{fig:tfidf_dists}, we clearly see that the results correspond to each other: $C_{hard}$ and $C_{elec}$ contain a more diverse genre spectrum (e.g., ``hardrock'' and ``hiphop'' are both part of $C_{hard}$'s top genres) than $C_{folk}$ and $C_{ambi}$ (e.g., in $C_{ambi}$'s top genres, we find ``ambient'' and ``darkambient''). 

%%%%%%%%%%%%%%%%%%%%%%%%%%%%%%%%%%%%%%%%%%%%%%
\subsubsection{Acoustic Feature Distributions}
\begin{figure}[t!]
    \centering
    \includegraphics[width=\textwidth]{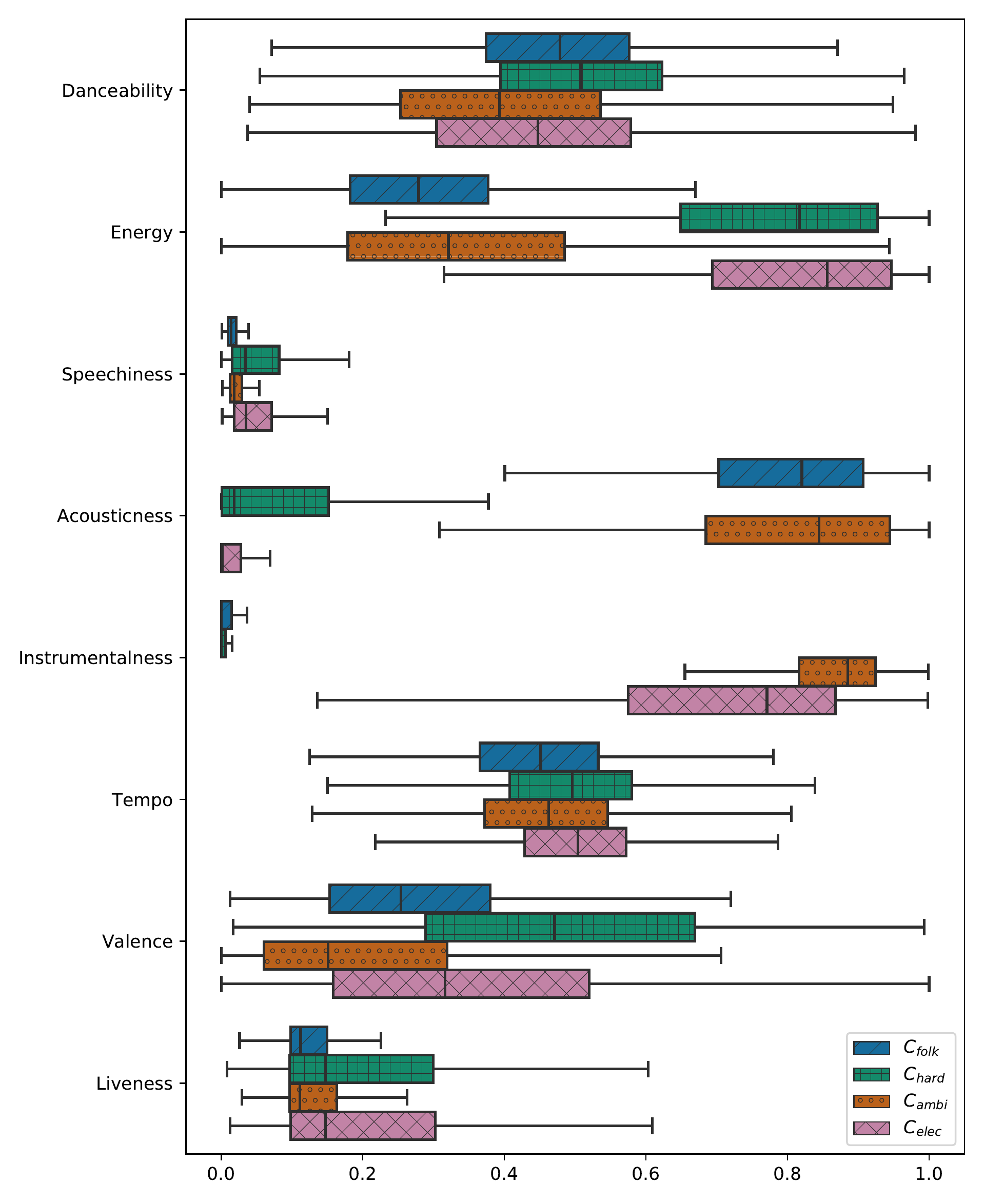}
    \caption{Distribution of the eight acoustic features for the four music clusters. While the clusters do not show obvious differences with respect to danceability and tempo, we find large differences with respect to energy, acousticness and instrumentalness.}
    \label{fig:acoustic_features}
\end{figure}

To understand the musical content of these four music clusters, we analyze the acoustic feature distributions of the four music clusters using boxplots in Figure~\ref{fig:acoustic_features}. This visualization does not show any obvious differences with respect to danceability and tempo among the four clusters. For the acoustic features energy, speechiness, acousticness, valence, and liveness, there are similar values for the cluster pairs \rev{$C_{folk}$ and $C_{ambi}$, and $C_{hard}$ and $C_{elec}$}. We observe differences between these two cluster pairs with respect to energy and acousticness. While \rev{$C_{hard}$ and $C_{elec}$ provide high energy values and small acousticness values, $C_{folk}$ and $C_{ambi}$} feature small energy values and high acousticness values.

In contrast, for instrumentalness, we see similar values for the cluster pairs \rev{$C_{folk}$ and $C_{hard}$ as well as for $C_{ambi}$ and $C_{elec}$. We observe very high values for $C_{ambi}$ and $C_{elec}$, and very small values for $C_{folk}$ and $C_{hard}$}. This difference is also visible in Figure~\ref{fig:track_cluster} in the form of the gap between \rev{$C_{folk}$ and $C_{hard}$ on the left, and $C_{ambi}$ and $C_{elec}$} on the right.

Summing up, in \rev{$C_{folk}$, we find music with low energy, high acousticness, and low instrumentalness; $C_{hard}$ contains music with high energy, low acousticness, and low instrumentalness; in $C_{ambi}$, we observe music with low energy, high acousticness, and high instrumentalness; and in $C_{elec}$}, we find high energy, low acousticness, and high instrumentalness. 
\rev{Thus, these findings are in line with the genre distributions presented in Figure~\ref{fig:tfidf_dists}.} 

%%%%%%%%%%%%%%%%%%%%%%%%%%%%%%%%%%%%%%%%%%%%%%%%%%%%%%%%%%%%%%%%%%%
\subsection{Assigning \rev{and Studying Beyond-Mainstream Music Listeners}}
\label{subsec:user_clustering}
In the next step, we assign the 2,074 \emph{BeyMS} users to the four music clusters to categorize them into four distinct beyond-mainstream subgroups for further analyses.

For each user $u$, we count the number of listening events $LE_{u,c}$ that $u$ has contributed to the tracks in each cluster $c$, where $c \in C = \{ C_{folk}, C_{hard}, C_{ambi}, C_{elec} \}$.
Then, we assign $u$ to the cluster $c$ for which the number of contributed listening events $LE_{u,c}$ is the highest.
However, because we have varying cluster sizes, the probability of $u$ listening to a track $t$ of the two larger clusters $C_{hard}$ and $C_{elec}$ is much higher than for the two smaller clusters $C_{folk}$ and $C_{ambi}$, although $C_{folk}$ and $C_{ambi}$ could be more representative choices for $u$. 
Thus, similar to the IDF distribution of genres (see Figure~\ref{fig:genre_IDF}), we take advantage of the IDF scoring to reduce the influence of the larger clusters and to assign higher weights to the smaller clusters. 
Specifically, these cluster IDF-scores are given by $IDF(c) = \log_{10}\frac{|T|}{|\{t \in T \mathrm{\,with\,} c_t\}|}$, 
i.e., by relating the number of all tracks $|T|$ to the number of tracks in cluster $c$ where $c_t$ is the music cluster assigned to track $t$.  
That lets us define the user--cluster weight $w_{u,c}$ 
for user $u$ and cluster $c$ as $w_{u,c} = IDF(c) \cdot LE_{u,c}$.

Consequently, users are assigned to the highest weighted music cluster and thus, a subgroup $U_c$ for cluster $c$ is given by $U_c = \{ u \in U: \argmax_{ c \in C}(w_{u,c}) \}$.

Out of the 2,074 $BeyMS$ users, we can assign 2,073 users to these subgroups. Thus, only 1 user listened to tracks not contained in any cluster in Figure~\ref{fig:track_cluster}.
Similar to the naming scheme of music clusters, we label the subgroups according to the name of their assigned music cluster. Hence, we obtain four subgroups $U_{folk}$, $U_{hard}$, $U_{ambi}$, and $U_{elec}$.

\begin{table}[t!]
    \centering
    \caption{Descriptive statistics of the four subgroups. Here, $|U|$ is the number of users, $|A|$ is the number of artists, $|T|$ is the number of tracks, $|LE|$ is the number of listening events, $|G|$ is the number of genres, $\overline{|LE_u|}$ is the average number of listening events per user, $\overline{|T_u|}$ is the average number of tracks per user and $\overline{Age}$ is the average age (along with the standard deviation) of users in the group.}
    \resizebox{\textwidth}{!}{
    \begin{tabular}{l| r r r r r| r r| r}
        \toprule
        Subgroup & $|U|$ & $|A|$ & $|T|$ & $|LE|$ & $|G|$  & $\overline{|LE_u|}$ & $\overline{|T_u|}$ & $\overline{Age}$ (std.)  \\ \midrule
        $U_{folk}$ & 369 & 9,559 & 72,663 & 702,635 & 811 & 1,904.160 & 549.650 & 27.599 ($\pm$ 10.369)\\
        $U_{hard}$ & 919 & 11,966 & 107,952 & 2,150,246 & 1,274 & 2,339.767 & 557.470 & 23.867 ($\pm$ 8.912)\\
        $U_{ambi}$ & 143 & 6,869 & 39,649 & 224.327  & 918 & 1,568.720 & 473.308 & 29.571 ($\pm$ 14.138)\\
        $U_{elec}$ & 642 & 11,814 & 105,907 & 1,416,354  & 1,005 & 2,206.159 & 670.402 & 24.639 ($\pm$ 7.886) \\ \bottomrule
    \end{tabular}}
    \label{tab:user_group_stats}
\end{table}

Table~\ref{tab:user_group_stats} provides basic descriptive statistics of these four resulting subgroups. 
Here, $U_{hard}$ is the largest subgroup with $|U|$ = 919 users, followed by $U_{elec}$ with $|U|$ = 642 users, $U_{folk}$ with $|U|$ = 369 users, and $U_{ambi}$ with $|U|$ = 143 users. The differences with respect to the number of users also correspond to the differences regarding the number of artists $|A|$, the number of tracks $|T|$, and the number of listening events $|LE|$ contained in the clusters. In the case of the number of genres~$|G|$, this differs slightly because the users in the smaller $U_{ambi}$ cluster listen to more genres (i.e., 918) than the bigger $U_{folk}$ cluster (i.e., 811). This indicates that the users in $U_{ambi}$ listen to a broader set of music than the users in $U_{folk}$.

Considering the average number of listening events per user (i.e., $\overline{|LE_u|}$) and the average number of tracks per user (i.e., $\overline{|T_u|}$), we see that, while there is little difference between $U_{hard}$ and $U_{elec}$ with respect to $\overline{|LE_u|}$, $\overline{|T_u|}$ is much higher for $U_{elec}$ (i.e., 670.402) than for $U_{hard}$ (i.e., 557.470). This indicates that, although the number of listening events is nearly the same, users of $U_{elec}$ tend to listen to a wider set of tracks than users of $U_{hard}$. With respect to the average age of the users $\overline{Age}$, we see that the users of $U_{folk}$ and $U_{ambi}$ are the oldest ones, and users of $U_{hard}$ and $U_{elec}$ are the youngest ones. 
However, it is worth noting that the group with the highest average age (i.e., $U_{ambi}$) also shows by far the highest standard deviation of age (i.e., 14.138 years).

\begin{figure}[t!]
    \centering
    \includegraphics[width=0.8\textwidth]{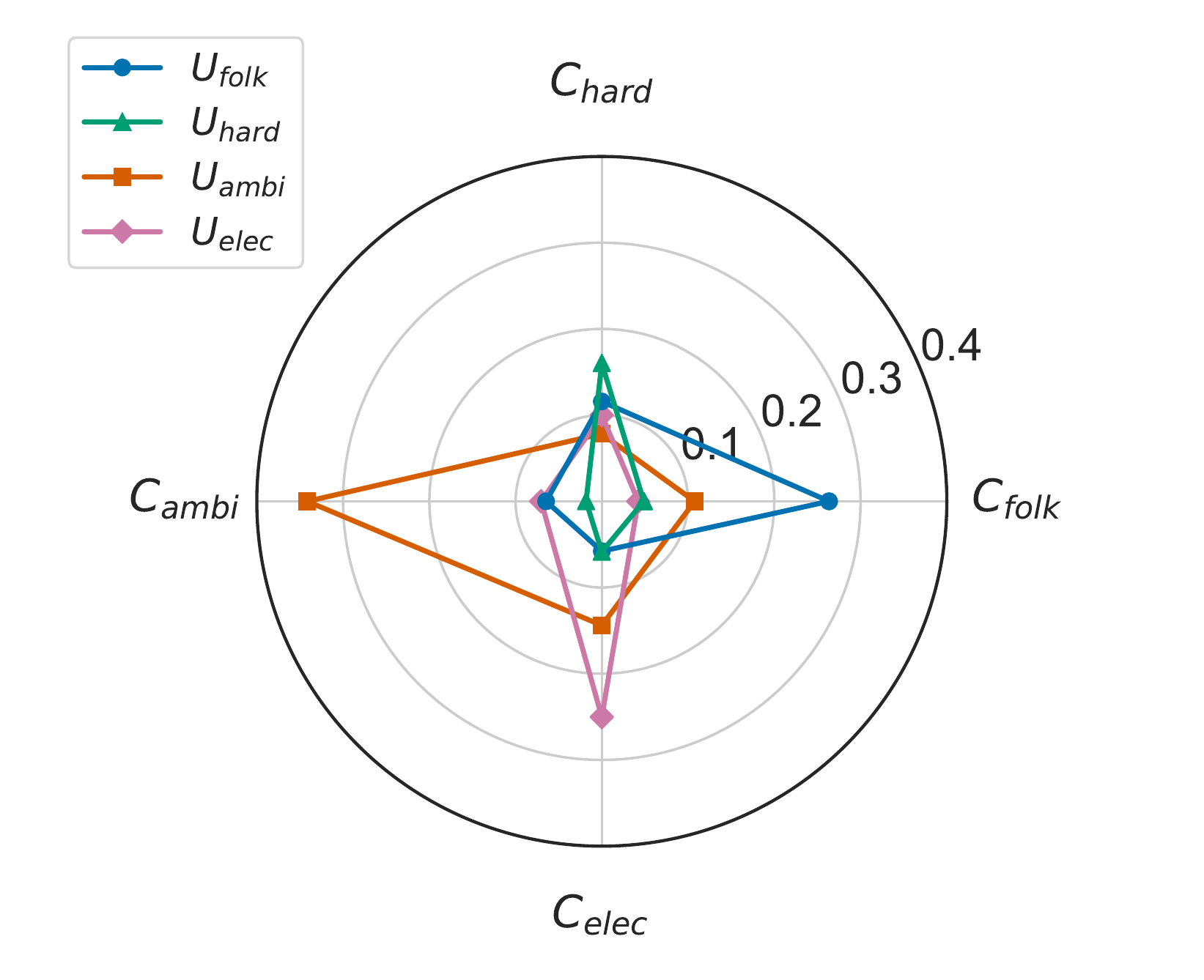}
    \caption{Radar plot illustrating the contribution of each music cluster to a subgroup. While the weight distribution of $U_{hard}$ and $U_{elec}$ is rather narrow, it is more broad in case of  $U_{folk}$ and $U_{ambi}$ suggesting that these groups are more open to music outside the own music cluster.}
    \label{fig:radarplot}
\end{figure}

In Figure~\ref{fig:radarplot}, we show the contribution of each music cluster to each subgroup in the form of a radar plot.
For this, we use the user-cluster weights $w_{u,c}$ introduced before and calculate the average weight over all users in cluster $c$. 
\rev{One consequence of the IDF scoring applied to $w_{u,c}$ is that the weight contributions of a user group to the four clusters does not sum up to 1, which eventually influences the interpretation of the values shown in Figure~\ref{fig:radarplot}. However, in return, these values account for the varying cluster sizes and can also be interpreted as preference weights for a user group towards a specific music cluster.} 

We observe that the weight distribution of the two larger subgroups $U_{hard}$ and $U_{elec}$ is rather narrow, which indicates that these users do not listen to many tracks of other clusters.
Contrary to that, the weights of the two smaller subgroups $U_{folk}$ and $U_{ambi}$ are more broadly distributed over the four music clusters. 
This suggests that users of $U_{folk}$ and $U_{ambi}$ are more open to music outside of their own music cluster than users of $U_{hard}$ and $U_{elec}$. 

%%%%%%%%%%%%%%%%%%%%%%%%%%%%%%%%%%%%%%%%%%%%%%%%%%%%%%
\subsubsection{Correlation of Music Clusters and Beyond-Mainstream Subgroups}
\begin{figure}[t!]
    \centering
    \includegraphics[width=0.8\textwidth]{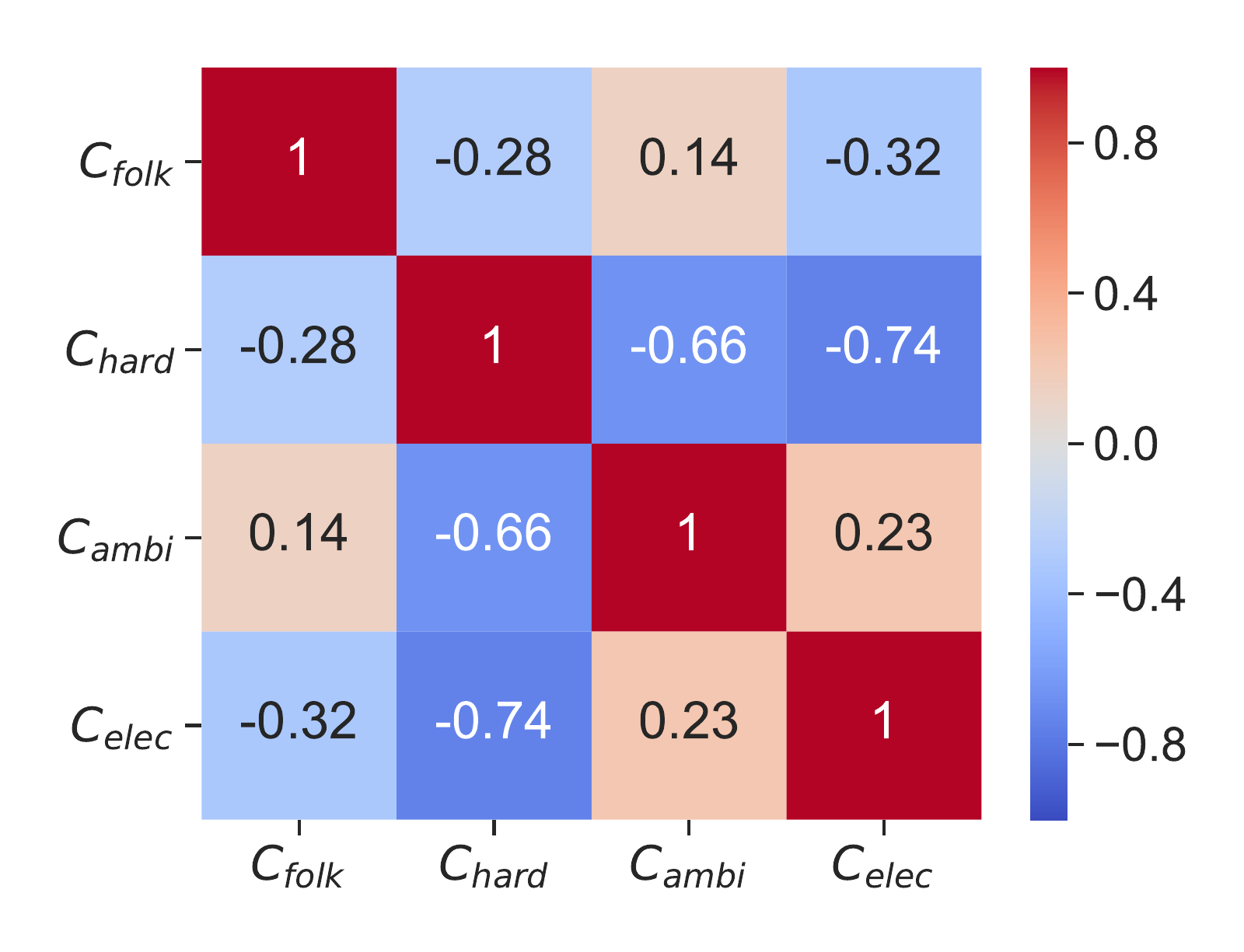}
    \caption{Pearson correlation matrix of the four music clusters. While $C_{hard}$ has solely negative correlations with all other clusters, and thus, listeners of $C_{hard}$ seem to be the most closed subgroup, $C_{ambi}$ has positive correlations with $C_{folk}$ and $C_{elec}$, and thus, listeners of $C_{ambi}$ seem to be the most open subgroup.}
    \label{fig:correlation}
\end{figure}

To better understand the correlations and connections between the music clusters and subgroups, we plot the Pearson correlation matrix of the four music clusters as a heatmap in Figure~\ref{fig:correlation}.
Here, we represent each music cluster $c$ by a 2,073-dimensional vector (i.e., one entry for each user) consisting of the user--cluster weights $w_{u,c}$, introduced before. Each element in the matrix is then calculated using the Pearson correlation measure based on these cluster vectors. 
For example, if there is a positive correlation between two clusters, we assume that a user who enjoys music from the one cluster likely also enjoys music from the other cluster. 
This can give us also an indication of the openness of a subgroup for music mainly listened to by other subgroups. 
Specifically, for $C_{folk}$, we see a positive correlation between $C_{folk}$ and $C_{ambi}$, and a negative correlation between $C_{folk}$ and both, $C_{hard}$ as well as $C_{elec}$.
Users listening to the music of $C_{hard}$ seem to represent the most closed subgroup as $C_{hard}$ because it solely has negative correlations with all other clusters, especially with $C_{ambi}$ and $C_{elec}$.
In contrast, users listening to the music of $C_{ambi}$ seem to represent the most open subgroup as $C_{ambi}$ has positive correlations with two other clusters, i.e., $C_{folk}$ and $C_{elec}$.
The fourth cluster, $C_{elec}$, is negatively correlated with $C_{folk}$ and especially with $C_{hard}$, 
and positively correlated with $C_{ambi}$.  
These results are also in line with the ones shown in Figure~\ref{fig:radarplot}, in which we identify the users of $U_{ambi}$ as more open music listeners than the ones of $U_{hard}$.

\begin{figure}[t!]
    \centering
    \includegraphics[width=0.8\textwidth]{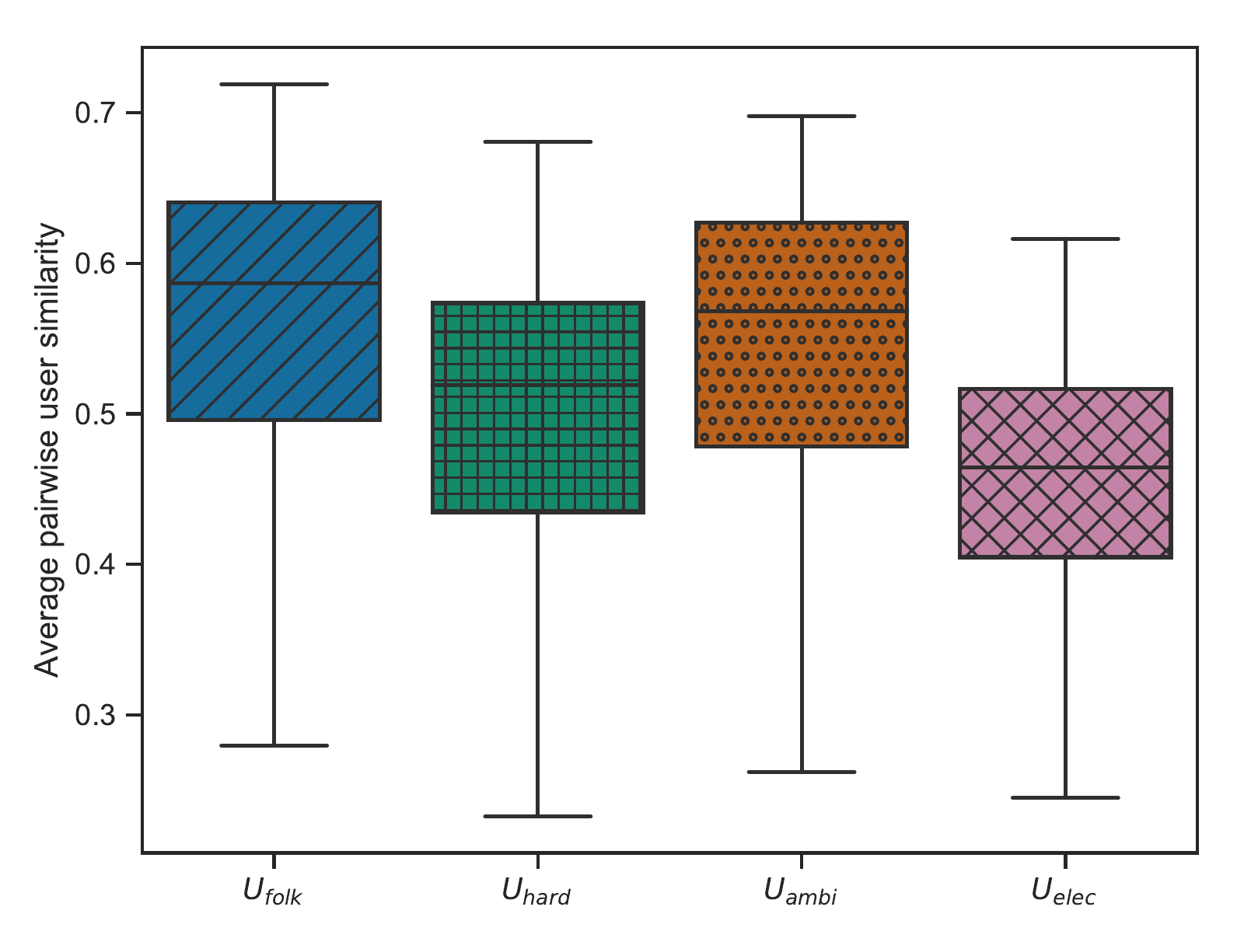}
    \caption{Boxplots showing the average pairwise user similarity of the four subgroups using the cosine similarity calculated on the users' genre distributions. While the users in $U_{hard}$ and $U_{elec}$ exhibit a more diverse listening behavior, users in $U_{folk}$ and $U_{ambi}$ tend to listen to more similar, i.e., less diverse, music genres.}
    \label{fig:similarity}
\end{figure}

In order to relate the openness of the subgroups to the diversity of the users within the subgroups, we calculate the average pairwise user similarity using the cosine similarity metric computed on the users' genre distributions, i.e., number of listening events per genre. Figure~\ref{fig:similarity} shows the resulting boxplots for the four identified subgroups (i.e., $C_{folk}$, $C_{hard}$, $C_{ambi}$, and $C_{elec}$). 
Figure~\ref{fig:similarity} shows that 
users in $U_{hard}$ and $U_{elec}$ have a rather small average pairwise user similarity and, thus, exhibit a more diverse listening behavior, whereas users in $U_{folk}$ and $U_{ambi}$ tend to listen to more similar music genres and, thus, have a 
narrow listening behavior within the group. 
Summed up, we find pronounced differences with respect to openness and diversity across the subgroups. Although $U_{ambi}$ is the most open subgroup (i.e., also listens to music of other subgroups), it is also the least diverse subgroup (i.e., the users within the group listen to very similar music). That observation is in line with what is shown in Figures~\ref{fig:tfidf_dists}, and Figure~\ref{fig:top10genres}. Here, we see that $C_{ambi}$, i.e., the most tightly connected music cluster to $U_{ambi}$, contains the dominating genre ``ambient'' as well as genres that are strongly associated with this dominating genre (e.g., ``darkambient''). 
For $U_{hard}$, we observe the opposite. While it is the least open subgroup, it is also the most diverse one (e.g., it contains ``hardrock'' as well as ``hiphop'' listeners).

%%%%%%%%%%%%%%%%%%%%%%%%%%%%%%%%%%%%%%%%%%%%%%%%%
\subsubsection{Recommendations for Beyond-Mainstream \rev{User} Subgroups} \label{s:recs_usergroups}
In Section~\ref{s:rec_first}, we have shown that the recommendation accuracy of four personalized recommendation algorithms is significantly worse for \emph{BeyMS} users than for \emph{MS} users. Now, we extend this analysis and evaluate the recommendation accuracy of these algorithms for the four subgroups (i.e., $U_{folk}$, $U_{hard}$, $U_{ambi}$, and $U_{elec}$).

Table~\ref{tab:mae_user_groups} shows our results with respect to the mean absolute error (MAE). Additionally, we analyze these results with respect to statistically significant differences in Table~\ref{tab:user_group_diffs} by performing ANOVA ($\alpha = .01$) and a subsequent Tukey-HSD test ($\alpha = .05$).
Here, we report pairwise differences as significant (marked with **), if both ANOVA and Tukey-HSD were significant across all five folds (see Section~\ref{s:rec_first} for details on the experimental setup). 

We see that among all algorithms, the significantly worst accuracy results (i.e., the highest MAE scores) are achieved for the $U_{hard}$ subgroup. Next, $U_{folk}$, $U_{ambi}$ and $U_{elec}$ reach significantly better (i.e., lower MAE scores) than $U_{hard}$ for all algorithms. However, there is no statistically significant difference between the recommendation accuracy of $U_{folk}$ and $U_{elec}$. The overall best accuracy results (i.e., lowest MAE scores) are reached for the $U_{ambi}$ subgroup. These results are also statistically significant when compared with the other subgroups for the NMF algorithm. NMF also gives the overall best accuracy results for all subgroups, which is in line with our results presented in Section~\ref{s:rec_first} \rev{and in our previous work~\cite{kowald2020unfairness}}.

Furthermore, we find a relationship between openness, diversity, and recommendation quality. Here, $U_{hard}$ is the least open but most diverse subgroup and gets the worst recommendations, while $U_{ambi}$ is the most open but least diverse subgroup and gets the best recommendations. 
This is in line with the findings of~\cite{tintarev2013adapting}, who have shown that users are more likely to accept recommendations from different groups (i.e., openness) rather than varied within a group (i.e., diversity). 
\rev{Thus, we find a relationship between the quality of recommendations provided to beyond-mainstream music listeners and openness as well as diversity patterns of these users.} 

\begin{table}[t!]
    \centering
        \caption{Mean absolute error (MAE) measurements for the four subgroups and four personalized recommendation algorithms. NMF (in bold) outperforms all other algorithms for all subgroups. Among the subgroups, the best accuracy results (i.e., lowest MAE scores) are reached by $U_{ambi}$, while the worst accuracy results (i.e., highest MAE scores) are reached by $U_{hard}$. 
        To facilitate comparison, we also show the MAE measurements for the \emph{BeyMS} and \emph{MS} user groups.} 
    \begin{tabular}{l | c c c c}
        \toprule
        Subgroup & UserItemAvg & UserKNN & UserKNNAvg & NMF \\ \midrule
        $U_{folk}$ & 63.2143 & 70.3049 & 67.4406 & \textbf{57.2278} \\
        $U_{hard}$ & 65.1464 & 73.1949 & 69.2855 & \textbf{59.6887} \\
        $U_{ambi}$ & 60.5558 & 69.8315 & 65.5708 & \textbf{54.2073} \\
        $U_{elec}$ & 62.2894 & 71.0387 & 66.1499 & \textbf{56.6209} \\\midrule
        \emph{BeyMS} & 63.4608 & 71.6694 & 67.5856 & \textbf{57.7703} \\
        \emph{MS} & 61.2562 & 68.4894 & 63.3985 & \textbf{54.8182} \\
        \bottomrule
    \end{tabular}
    \label{tab:mae_user_groups}
\end{table}

\begin{table}[t!]
    \centering
    \caption{Statistically significant differences between pairs of subgroups, as determined by ANOVA ($\alpha = .01$) and a subsequent Tukey-HSD test ($\alpha = .05$).} 
    \resizebox{\textwidth}{!}{
    \begin{tabular}{l | c c c c | c c c c | c c c c | c c c c}
        \toprule
        & \multicolumn{4}{c|}{UserItemAvg} & \multicolumn{4}{|c|}{UserKNN} & \multicolumn{4}{|c|}{UserKNNAvg} & \multicolumn{4}{|c}{NMF} \\ \midrule
        Subgroup & $U_{folk}$ & $U_{hard}$ & $U_{ambi}$ & $U_{elec}$ & $U_{folk}$ & $U_{hard}$ & $U_{ambi}$ & $U_{elec}$ & $U_{folk}$ & $U_{hard}$ & $U_{ambi}$ & $U_{elec}$ & $U_{folk}$ & $U_{hard}$ & $U_{ambi}$ & $U_{elec}$ \\ \midrule
        $U_{folk}$ &  & ** & ** &  &    & ** &  &  &    & ** &  &  &    & ** & ** &  \\
        $U_{hard}$ & ** &  & ** & ** &  **  &  & ** & ** &  **  &  & ** & ** &  **  &  & ** & ** \\
        $U_{ambi}$ & ** & ** &  &  &    & ** &  &  &    & ** &  &  &  **  & ** &  & ** \\
        $U_{elec}$ &  & ** &  &  &    & ** &  &  &    & ** &  &  &    & ** & ** &  \\ \bottomrule
    \end{tabular}}
    \label{tab:user_group_diffs}
\end{table}

\begin{figure}[t!]
    \centering
    \includegraphics[width=0.8\textwidth]{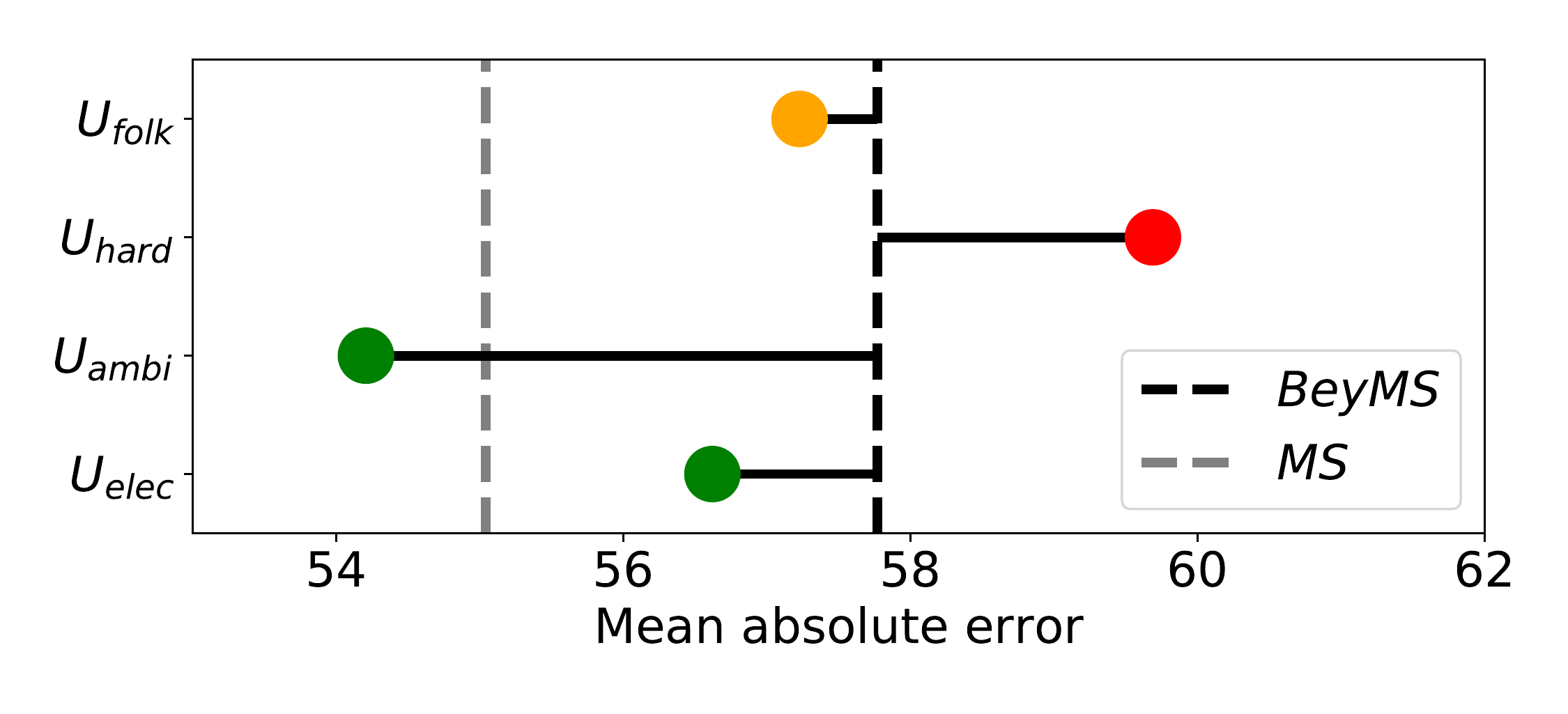}
    \caption{Comparison of the mean absolute error (MAE) scores reached by NMF for the four subgroups with the ones reached by NMF for \emph{BeyMS} (black dashed line) and \emph{MS} (grey dashed line). While specific subgroups (i.e., $U_{hard}$) are treated in an unfair way by recommendation algorithms, others (i.e., $U_{ambi}$) are not.}
    \label{fig:mae_user_groups}
    \end{figure}

Finally, in Figure~\ref{fig:mae_user_groups}, we visually compare the MAE scores reached by \rev{the best performing approach} NMF for the four subgroups. Additionally, we depict the MAE score for \emph{BeyMS} as a black dashed line and the MAE score for \emph{MS} as a grey dashed line. We see that $U_{hard}$ reaches worse results than \emph{BeyMS} while $U_{folk}$ and $U_{elec}$ reach slightly better results than \emph{BeyMS}.
Interestingly, $U_{ambi}$ not only reaches better results than \emph{BeyMS} but also better results than \emph{MS}. Although this improvement over \emph{MS} is not statistically significant (according to a one-tailed Mann-Whitney-U test with $\alpha = .0001$), it shows that there is a large variety among \emph{BeyMS} users, where specific subgroups (i.e., $U_{hard}$) are disadvantaged in terms of recommendation accuracy by recommendation algorithms while others (i.e., $U_{ambi}$) are not. 

\section{Conclusions and Future Work} \label{s:conc}
\rev{In this paper, we shed light on the characteristics of beyond-mainstream music and music listeners.} 
As our first contribution, we identified 2,074 beyond-mainstream music listeners (i.e., \emph{BeyMS}) in the Last.fm platform, and subsequently created a novel dataset \rev{called} \emph{LFM-BeyMS} based on the listening histories of these users. We further enriched this dataset with (i)~acoustic features of music tracks gathered from Spotify, and (ii)~genre information of tracks derived from Last.fm tags and matched with the Spotify microgenre taxonomy. Additionally, \rev{for reasons of comparability,} \emph{LFM-BeyMS} contains data of 2,074 Last.fm users listening to mainstream music. Using this dataset, as our second contribution, we \rev{validated related research by showing} that beyond-mainstream music listeners receive a significantly lower recommendation accuracy than mainstream music listeners by four standard recommendation algorithms (i.e., UserItemAvg, UserKNN, UserKNNAvg and NMF). 

As our third contribution, we applied the clustering algorithm HDBSCAN* on the acoustic features of tracks listened by \emph{BeyMS} and identified four clusters of beyond-mainstream music\rev{: (i)~$C_{folk}$, music with high acousticness such as ``folk'', (ii)~$C_{hard}$, high energy music such as ``hardrock'', (iii)~$C_{ambi}$, music with high acousticness and instrumentalness such as ``ambient'', and (iv)~$C_{elec}$, music with high energy and instrumentalness such as ``electronica''.}

\rev{As our fourth contribution,} we mapped these clusters to our \emph{BeyMS} users, which led to \rev{four beyond-mainstream subgroups: (i)~$U_{folk}$, (ii)~$U_{hard}$, (iii)~$U_{ambi}$, and (iv)~$U_{elec}$.} We analyzed these subgroups with respect to their openness (i.e., across-groups diversity -- do users of one group listen to music of other groups?) and diversity (i.e., within-groups diversity -- how dissimilar is the music listened to by users within groups?). Here, we found large differences between $U_{hard}$ and $U_{ambi}$. Although $U_{hard}$ is the most closed subgroup (i.e., users do not listen to music of other subgroups), it is also the most diverse subgroup (i.e., users listen to a diverse set of genres such as ``hardrock'' and ``hiphop''). For $U_{ambi}$, we get opposite results: while it is the most open subgroup (i.e., users listen to music of other subgroups as well), it is also the least diverse one (i.e., the users within the group listen to very similar music such as ``ambient'' and ``darkambient''). We related these characteristics of the subgroups to the recommendation quality of the four recommendation algorithms UserItemAvg, UserKNN, UserKNNAvg and NMF. Here, we found that $U_{hard}$ got music recommendations with lowest accuracy, while $U_{ambi}$ got music recommendations with highest accuracy.
\rev{This is in line with related research~\cite{tintarev2013adapting}, which has shown that openness is stronger correlated with accurate recommendations than diversity.} 
$U_{ambi}$ even received better recommendations than the \rev{group of} mainstream music listeners. This result highlights that there are large differences between the subgroups of beyond-music listeners. \rev{Finally, to foster reproducibility of our research, we provide our novel \emph{LFM-BeyMS} dataset via Zenodo as well as our source code via Github.}

We believe that our findings provide useful insights for creating user models and recommendation algorithms that better serve beyond-mainstream music listeners. 
As it was shown in~\cite{kowald2020unfairness}, beyond-mainstream music listeners tend to have larger user profile sizes than users interested in mainstream music, which means that they provide a substantial amount of listening interaction data for services such as Last.fm and Spotify. We assume that improving the recommendation quality for this active user group also leads to another effect, namely a more prominent exposure of (long-tail) music artists due to a better-connected recommendation network~\cite{lamprecht2017method}. We leave such investigations to future work. 

\paragraph{Limitations and future work.}
\rev{Despite the merits of this work, we are aware of its limitations.}
The first limitation we recognize is that our analyses are based on a sample of the Last.fm community. The extent to which their listening behavior is representative of the Last.fm community at large, or similar music streaming communities such as Spotify, needs further investigation.

Next, since we conducted a comparative study of the accuracy of recommender systems algorithms---and were therefore not interested to beat state-of-the-art algorithms---we focused on traditional algorithms (e.g., KNN-based collaborative filtering) instead of investigating the most current deep learning architectures, which would also require a much higher computational effort. Furthermore, an award-winning-paper by Dacrema et al.~\cite{DBLP:conf/recsys/DacremaCJ19} has recently shown that traditional algorithms are able to outperform almost all deep learning architectures.

While our work serves as a first milestone towards better characterizing beyond-mainstream \rev{music and listeners of such music}, future work should focus on user modeling techniques to individually target the different subgroups, for example by integrating knowledge about openness and diversity. With respect to analyzing openness and diversity of users and user groups, we would also like to work on a more formal definition of these dimensions, which would not only allow us to measure them more precisely but also to integrate them into the recommendation calculation process.

\rev{Additionally, since previous research has shown that the listener's cultural background impacts the quality of music recommendations~\cite{zangerle:tismir20}, we plan to compare the cultural and socioeconomic aspects of beyond-mainstream and mainstream music listeners. 
We plan to employ these aspects by means of Hofstede's cultural dimensions~\cite{hofstede2010cultures} and the World Happiness Report~\cite{helliwell2013world}.}

Finally, another avenue for future work is the research in the area of fair music recommender systems. Here, we plan to build user models that are capable of accounting for the complex characteristics of beyond-mainstream music listeners presented in this paper. While we believe that more specialized user models could help to provide better recommendations for users who currently receive worse recommendations (e.g., the $U_{hard}$ subgroup identified in this paper), we also aim to highlight that such user models still need to be generalizable to avoid any unfair treatment of other users. Hence, future research should work on achieving a specialization-generalization trade-off in music recommender systems. We hope that our open \emph{LFM-BeyMS} dataset \rev{as well as our source code} will be of use to the scientific community for subsequent analyses.

%\end{linenumbers}

%%%%%%%%%%%%%%%%%%%%%%%%%%%%%%%%%%%%%%%%%%%%%%%%%%%%%%%%%%%%%%%%%%%%%%%%%%%%%%%%

\begin{backmatter}

\section*{Availability of Data and Materials}
The \rev{\emph{CultMRS}} dataset can be found on Zenodo~\url{https://doi.org/10.5281/zenodo.3477842}. Additionally, we provide our novel \emph{LFM-BeyMS} dataset via Zenodo:~\url{https://doi.org/10.5281/zenodo.3784764}. 
Our Python-based implementations are available via Github~\url{https://github.com/pmuellner/supporttheunderground}. 

\section*{Competing interests}
The authors declare that they have no competing interests.

\section*{Funding}
This work is funded by the TU Graz Open Access Publishing Fund and the Austrian Science Fund (FWF): V579.

\section*{Author's contributions}
All authors contributed to manuscript revision, read, and approved the submitted version.

\section*{Acknowledgements}
This work is supported by the Know-Center GmbH within the Austria FFG COMET program.

\printendnotes

% if your bibliography is in bibtex format, use those commands:
\bibliographystyle{bmc-mathphys} % Style BST file (bmc-mathphys, vancouver, spbasic).
\bibliography{article}      % Bibliography file (usually '*.bib' )

\end{backmatter}
\end{document}